\documentclass{aa}  

\usepackage{graphicx}
\usepackage{lscape}
\usepackage{txfonts}
\usepackage{xcolor}
\usepackage{tcolorbox}
\usepackage[colorinlistoftodos]{todonotes}
\usepackage{hyperref}
\usepackage{titlesec}
\usepackage{caption}
\usepackage{subcaption}
\usepackage{rotating}
\usepackage{longtable}
\usepackage{siunitx}

\urldef{\urlcal}\url{https://www.ph.unimelb.edu.au/~mtrenti/cvc/CosmicVariance.html}
\urldef{\urlhdfs}\url{https://archive.stsci.edu/pub/hdf_south/}
\urldef{\urlsof}\url{https://gitlab.com/SoFiA-Admin/SoFiA-2}
\urldef{\urlasc}\url{https://almascience.org/proposing/sensitivity-calculator}

\begin{document}

   \title{A deep ALMA Band~3 survey of HDFS/MUSE3D}
   \subtitle{Survey description and initial results}

   \author{Hugo Messias \inst{1,2}
          \and
          Laura Gomez \inst{1}
          \and
          Harold Francke \inst{1}
          \and
          Bill Dent \inst{1}
          \and
          Bel\'en Alcalde Pampliega \inst{2,3}
          \and
          Ruediger Kneissl \inst{1,2}
          \and
          Yiqing Song \inst{2}
          \and
          Dirk Petry\inst{4}
          \and
          Paulo Cort\'es \inst{1,5}
          \and
          Sergio Mart\'in \inst{1,2}
          }

   \institute{Joint ALMA Observatory, Alonso de C\'ordova 3107, Vitacura 763-0355, Santiago, Chile\\
              \email{hugo.messias@eso.org}
         \and
             European Southern Observatory, Alonso de C\'ordova 3107, Vitacura, Casilla 19001, Santiago de Chile, Chile\\
         \and
             Instituto de Estudios Astrofísicos, Facultad de Ingeniería y 455 Ciencias, Universidad Diego Portales, Av. Ejército Libertador 441, Santiago, Chile\\
         \and
             European Southern Observatory, Karl-Schwarzschild-Str. 2, 85748 Garching, Germany\\
         \and
             National Radio Astronomy Observatory, 520 Edgemont Road, Charlottesville, VA 22903, USA
             }

   \date{Received 31$^{\rm st}$ October 2024; accepted 20$^{\rm th}$ October 2025}
 
  \abstract
   {After more than 10 years of ALMA operations, the community interest in conducting deep, extra-galactic, millimetre surveys resulted in varying strategic compromises between areal size and map depth to survey the sky. The current bias leans towards a galaxy population found in the field or towards rich star-bursty proto-cluster groups, both tendentiously surveyed at coarse spatial resolutions.}
   {Here, we describe a survey addressing these biases. A deep 3\,mm survey was conducted with ALMA in long-baselines on a $1\times1\,$arcmin$^2$ region in the \textit{Hubble} Deep Field South (HDFS), also covered by the Multi Unit Spectroscopic Explorer (MUSE) in order to assess resolved molecular gas properties in galaxies in group environments at $z>1$.}
   {ALMA observations comprising a 4-pointing mosaic with a single Band~3 (3\,mm) spectral tuning were conducted to cover CO transitions from different groups identified by MUSE. This work consists in a total effective time on source of 61 hours in configurations with up to 15\,km baselines.}
   {The final data-set yields an angular resolution of $0.15\arcsec$--$0.2\arcsec$ (imaging weights dependent) and maximum recoverable scales of $1\arcsec$--$2\arcsec$. The final continuum map reaches an unprecedented sensitivity of {\sc rms}$\sim$2\,$\mu$Jy/beam, allowing the detection of three sources at 3\,mm (only one showing multi-wavelength counterparts from rest-frame UV to radio). Moreover, we detect six line emitters associated with CO~$J=2-1$ at $z_{spec}=1.284$, one of them previously undetected by MUSE and none detected in 3\,mm continuum. The inter-stellar medium gas masses range from $\sim2\times10^9$ to $\sim9\times10^{10}\,$M$_\odot$ (adopting $\alpha_{\rm CO}=4~{\rm M_\odot / (K.km/s.pc^2)}$, including Helium). Overall, this galaxy group is quite diverse with no two galaxies alike, some showing clear physical offsets with respect to \textit{Hubble} imaging tracing rest-frame ultra-violet emission. We also derive cosmic molecular gas mass densities using this sample as a reference for group environments, and we find that these yield comparable densities as the galaxy population found in field environments.}
   {}


   \maketitle

\section{Introduction}

The advent of the Atacama Large (sub-)Millimetre Array \citep[ALMA;][]{BrownWildCunningham04,cortes2025} enabled the community to conduct multiple deep extra-galactic surveys at unprecedented angular resolution and sensitivity with which to study inter-stellar medium (ISM) gas in galaxies in the early Universe \citep{Hatsukade16,Walter16,Dunlop17,Franco18,Hatsukade18,Hill24}. Different groups gave priority either to survey depth or size. Yet, ALMA brought another capability to the table that revolutionized our knowledge of the molecular gas content until cosmic noon and beyond ($z\gtrsim2$). By planning deep fields to be covered at multiple frequency setups ––– the so-called spectral scans --- and in different ALMA bands, the community was able to blindly detect the faint dust and molecular gas emissions from galaxy populations while determining at the same time their distance \citep[e.g.,][]{Walter16}. This resulted in exquisite galaxy samples with which to determine carbon monoxide (CO) luminosity functions, and, by extension, molecular gas (H$_2$) mass functions \citep{Decarli16,Decarli19,Riechers19,Lenkic20}. Despite being expected based on relations such as the Schmidt-Kennicutt law \citep{Schmidt59,Kennicutt98}, the community finally saw, for the first time, that the cosmological evolution of H$_2$ mass density ($\rho_{\rm H_2}$) is indeed closely followed by the star-formation rate density evolution \citep{Decarli16,Decarli19,Magnelli20,Walter20}, both peaking around 10\,Gyr ago ($z\sim2$; the so-called cosmic noon).

For practical reasons all these surveys, being detection experiments in nature, were conducted with compact configurations in order to avoid resolving out extended emission. As a result, the spatial resolutions of these surveys are planned to be of arcsec-scales. Moreover, being blind surveys within known extra-galactic fields, the underlying bias of these samples was toward field galaxy populations. However, it is now well known that group environment has a major role in the evolution of galaxies \citep{Fujita04,Vulcani15,Vulcani17,Bianconi18}. Yet, the surveys conducted by ALMA targeting galaxy groups are clearly biased toward rich proto-clusters \citep[e.g.,][]{Miller18,Oteo18}, where extremely star-bursting galaxies are found to be close-by ($<<1$\,Mpc). On the other hand, ALMA surveys towards peaks in the cosmic infrared background \citep[e.g.,][]{Kneissl19,Hill25} show also groups of main-sequence galaxies, with extreme star-forming rates \citep[see also][]{Miller18,Oteo18}. In order to change this trend, we conducted a $1\times1\,$arcmin$^2$ deep single-tuning survey at sub-arcsec scales with ALMA targeting two galaxy groups identified at $z_{\rm spec}=1.284$ and $4.699$ in a deep survey \citep[27\,h on source;][B15]{Bacon15} conducted with the Multi Unit Spectroscopic Explorer \citep[MUSE;][]{Bacon10} on the European Southern Observatory Very Large Telescope \citep{vltwb}.

The structure of this article is as follows: Section~\ref{sec:obs} describes the field selection, the choice of spatial and spectral coverages, the observational strategy, and the resulting survey properties; initial results are presented in Section~\ref{sec:results}, with a characterization of the survey depth, the achieved angular and maximum recovered scales, and a first characterization of the line and continuum emitters in the field. We present our concluding remarks in Section~\ref{sec:conc}.

Throughout this work, we adopt a flat cosmology with the following parameters\footnote{There are slight differences between values reported in the right-most column in Table~2 in \citet{Planck20} and the {\sc Astropy.Cosmology.Planck18} package. Specifically $\Omega_{\rm M}$, with reported values of 0.3111 and 0.30966, respectively, the difference is related to the parameter definition namely the inclusion or not of massive neutrinos (as reported in issue \#10957 in {\sc github}).}: $\rm{H_0} = 67.66$\,km~s$^{-1}$~Mpc$^{-1}$, $\Omega_{\Lambda} = 0.69$ and $\Omega_{\rm M} = 0.31$ \citep{Planck20}.

\section{Observations} \label{sec:obs}

\subsection{Field selection}

This work makes use of the ALMA Observatory Project\footnote{More information on observatory projects: https://almascience.eso.org/alma-data/observatory-projects} 2022.A.00034.S (PI B.~Dent) which conducted Band~3, long-baseline, extremely deep observations toward the \textit{Hubble} Deep Field South \citep[HDFS;][]{Ferguson00}. This field is historically aimed at studying galaxy group environments and comprises two patches of sky surveyed by the \textit{Hubble} Space Telescope (HST). One of these patches is toward a known quasar (J2233-6033 at $z_{\rm spec}=2.25$) while the other is a deep survey on a blank field West-ward of the quasar. This blank field was later covered by a MUSE-3D 1~arcmin$\times$1~arcmin footprint \citep{Bacon15}. It comprises 27\,h of on-source integration resulting in an emission-line surface brightness limit of $\sigma=1\times10^{-19}\,{\rm erg~s^{-1}~cm^{-2}~arcsec^{-2}}$ and a point-source emission line $5\sigma$ level of $3\times10^{-19} {\rm erg~s^{-1}~cm^{-2}}$ (within a 1\,arcsec aperture). These survey specifications allowed numerous galaxy groups to be identified at different epochs of the Universe \citep{Bacon15} and, as a result, these groups were considered potential targets of this ALMA survey.

\subsection{Choice of Spatial and Spectral coverage} \label{sec:specspatcov}

\begin{figure}
    \centering
    \includegraphics[width=0.48\textwidth]{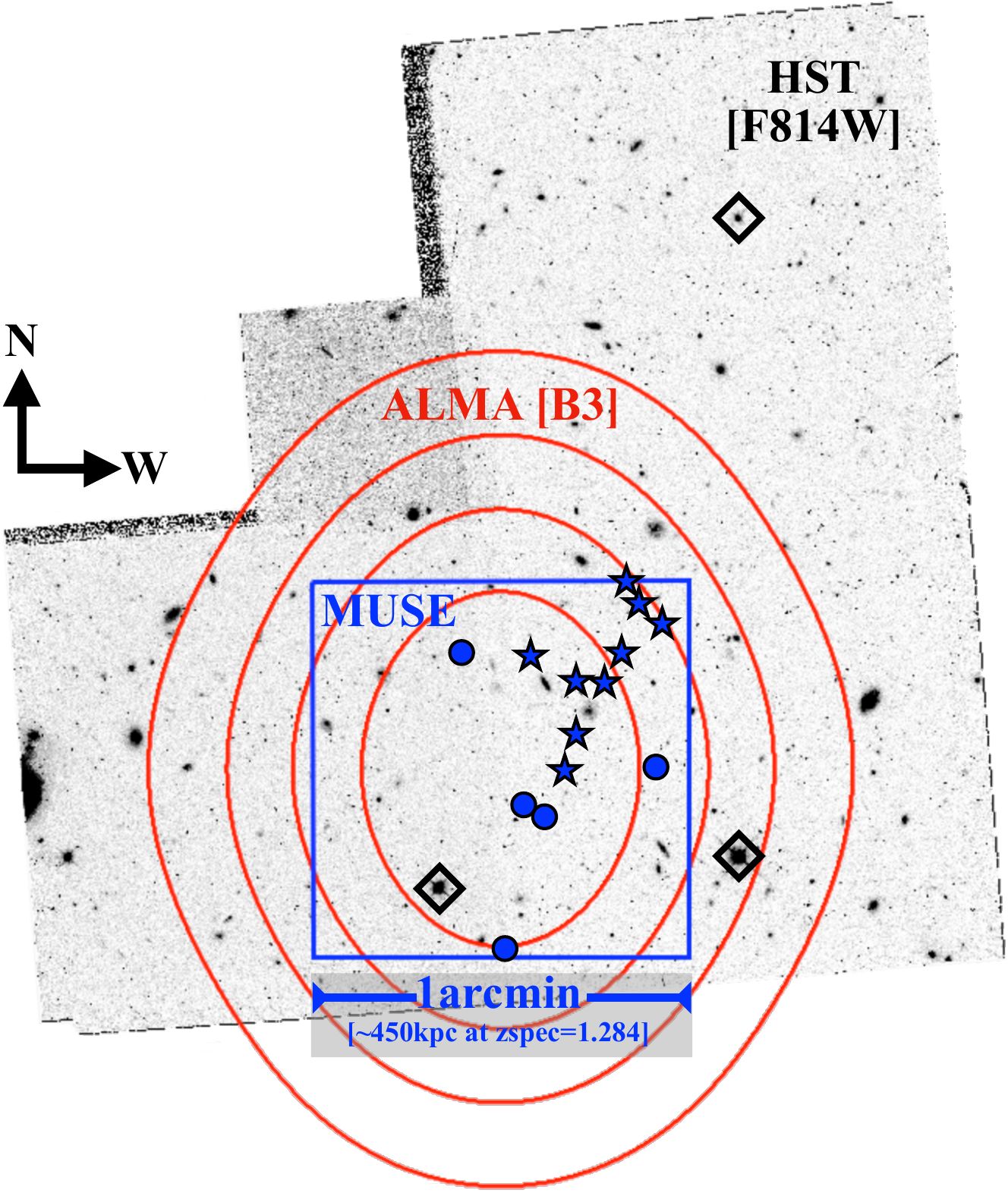}
    \caption{Comparison of the ALMA coverage (red iso-contours at primary beam attenuation levels of 0.8, 0.6, 0.4, and 0.2) with those of HDFS (WFPC2/\textit{F814W}; background grey map) and MUSE3D ($1\times1\,$arcmin$^2$ blue region). Blue stars and circles indicate the positions the members of the groups at $z_{\rm spec}=1.284$ and 4.699, respectively (note that two of the high-redshift members are too close to each other to distinguish their two circles). The HDFS imaging was aligned with \textit{Gaia}-DR3 reference \citep{Gaia23} making use of three stars in the field (black diamonds) at RA, Dec = 22:32:50.513, -60:34:00.94 (used for the MUSE Slow Guiding System); 22:32:56.999, -60:34:05.82 (brightest star in the field); and 22:32:50.513, -60:32:18.83. The image is oriented such that North is up, and West is right.}
    \label{fig:almaptg}
\end{figure}

In order to cover the 1~arcmin$\times$1~arcmin MUSE~3D survey region in ALMA Band~3, a mosaic comprising four pointings was observed. Figure~\ref{fig:almaptg} shows the ALMA coverage compared with those from the HDFS \citep[background grey map\footnote{\urlhdfs};][]{Ferguson00} and MUSE3D (blue region). Table~\ref{tab:almaptg} provides the absolute positions of the four observed pointings.

\begin{table}
\caption{Adopted ALMA mosaic pointing coordinates}             
\label{tab:almaptg}      
\centering          
\begin{tabular}{l r r} 
\hline\hline
Pointing & R.A. & Dec. \\
& [h:m:s] & [d:m:s] \\
\hline 
South & 22:32:55.64 & -60:34:12.5 \\
West & 22:32:53.64 & -60:33:47.0 \\
East & 22:32:57.64 & -60:33:47.0 \\
North & 22:32:55.64 & -60:33:21.5 \\
\hline
\end{tabular}
\end{table}

\begin{figure}
    \centering
    \includegraphics[width=0.48\textwidth]{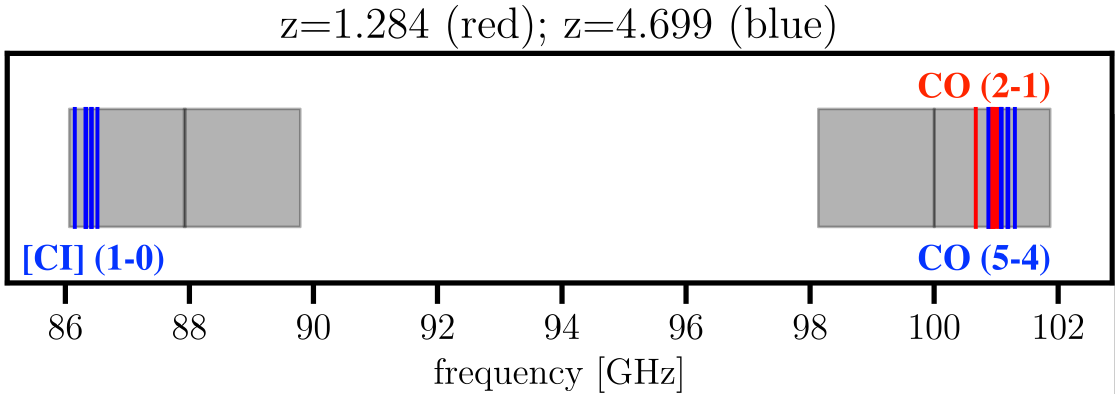}
    \caption{The spectral tuning used to conduct the survey was especially chosen to target the transitions CO~(2-1) and (5-4) towards the two galaxy groups identified by the MUSE~3D survey at $z=1.284$ and $4.699$, with 9 and 6 reported members, respectively (vertical lines correspond to each member). For the $z=4.699$ group, the [CI]~(1-0) transition is also covered.}
    \label{fig:tune}
\end{figure}

The ALMA spectral tuning was set to Frequency Division Mode (FDM) setup in Band\,3 with four 1.875\,GHz-wide Spectral Windows (SPWs; i.e, total continuum bandwidth 7.464\,GHz). The SPWs are centred at: 87.00, 88.86, 99.06, and 100.93\,GHz ($\nu_{\rm LO1}=93.96\,$GHz; $\lambda=3.2\,$mm). Each SPW has 240\,channels with an effective spectral resolution of 8.197\,MHz ($\sim24\,$km/s at 100.93\,GHz). Table~\ref{tab:grpcov} summarises how potential transitions from MUSE-identified galaxy groups are covered by the single spectral setup of our observations. The tuning was chosen such that CO transitions were optimally covered towards two galaxy groups identified by MUSE. Specifically, the groups identified by \citet{Bacon15} at $z=1.284$ (9 members) and 4.699 (6 members) have their CO\,(2-1) and (5-4) transitions, respectively, redshifted into the highest-frequency SPW in the data-set (Figure~\ref{fig:tune}). Moreover, [CI]\,($^3\mathrm{P}_1-{^3\mathrm{P}}_0$) at $z=4.699$ is redshifted into the lowest-frequency SPW. Nevertheless, other galaxy groups may also have different transitions covered by the other SPWs. However, the observation setup is not ideal for either of the two extra groups. Either the long-baseline configurations may be resolving out extended emission from the members of the $z_{\rm spec}=0.172$ group, or the systemic velocity uncertainty of the members within the $z_{\rm spec}=3.013$ group combined with the velocity spread of the group makes the CO transitions fall outside our spectral coverage. In the end, as one can see ahead, we could only identify CO emission from the members of the group at $z_{\rm spec}=1.284$.

\begin{table}
\caption{Covered galaxy groups identified by MUSE}             
\label{tab:grpcov}      
\centering          
\begin{tabular}{c c c c c c}
\hline\hline
$z$ & lines$_{\rm opt}$ & $v_{\rm rms}$ & ${\rm N_m}$ & CO & $\nu_{\rm obs}$ \\
& & [km/s] & & J$_{\rm up}$ & [GHz] \\
\hline 
0.172 & H$_\alpha$, [OIII], H$_\beta$ & 65 & 3 & 1\tablefootmark{a} & 98.354 \\
1.284 & [NeIII], [OII], MgII & 354 & 9 & 2 & 100.936 \\
3.013 & Ly$_{\alpha}$, S\,II, C\,IV, C\,II & 350 & 3 & 3\tablefootmark{b} & 86.169 \\
4.699\tablefootmark{c} & Ly$_{\alpha}$ & 430 & 6 & 5 & 101.117 \\
\hline
\end{tabular}
\tablefoot{The different groups identified by \citet{Bacon15} at different redshifts (\textit{z}; via different optical lines, lines$_{opt}$) for which the selected spectral tuning covers the expected observed frequency ($\nu_{\rm obs}$) of redshifted CO transitions (CO~J$_{\rm up}$). The number of constituents of each group is reported in column ${\rm N_m}$. Note that the systemic velocity of groups identified via Lyman-$\alpha$ is uncertain since this line tends to present significant offsets with respect to host galaxy systemic redshift \citep[e.g.,][]{Verhamme18}.
\tablefoottext{a}{The [CI]\,(1-0) emission from this group is also covered by the lowest-frequency SPW.}
}
\end{table}

\subsection{Observing Strategy and Calibration}

The observations were carried out from 2023-09-05 to 2023-11-30. During this period, the 12m array was re-configured covering ALMA configurations C-9 to C-7 \citep[see][for ALMA technical details]{cortes2022}. During the period between 2023-10-30 and 2023-11-27, the array was in a hybrid configuration, between C-8 and C-7, which we refer to as C-7hybrid (or C-7h) throughout the text. Table~\ref{tab:obssum} lists the number of executions in each distinct configuration and further detail is given in Appendix~\ref{app:ebs}.

\begin{table}
\caption{Summary of ALMA scheduling execution blocks (EBs) per configuration}             
\label{tab:obssum}      
\centering          
\begin{tabular}{l c} 
\hline\hline
Array & Number of \\
Config.& PASS EBs \\
\hline 
C-9 & 3 \\
C-8 & 52 \\
C-7h & 13 \\
C-7 & 2 \\
\hline
\end{tabular}
\end{table}

For our science target, the phase calibration was done with the quasar J2239-5701 (3.6\,deg away) with a cycle-time of 54\,s that was unchanged during the period within which the project was observed. The check source was J2207-5346 (5.4\,deg away from J2239-5701). Finally, in each execution, the bandpass and flux calibration made use of a single calibrator source (J2357-5311 or J2258-2758).

Each execution was expected to run for 1.8\,h ($\sim50\,$min on source). Overall, the total of 70 observing executions at ALMA resulted into 61\,h of effective time on-source, distributed over the 4 pointings. The data were calibrated with the ALMA Pipeline version 2022.2.0.68 \citep{Hunter23} and CASA 6.4.1.12\citep{casa2022}, which took the equivalent of 48\,days of single-cluster-node continuous computing time to process. Appendix~\ref{app:ebs} provides more detailed information on each of the considered EBs.

\subsection{Survey depth and noise properties} \label{sec:depth}

\begin{figure}
    \centering
    \includegraphics[width=0.52\textwidth]{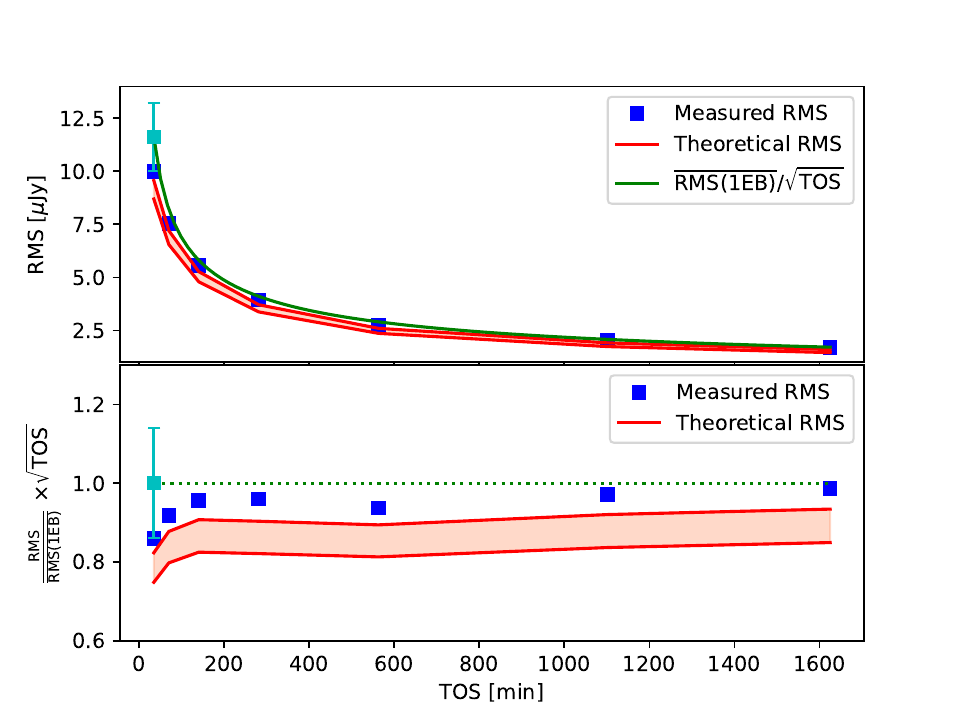}
    \caption{Sensitivity (noise {\sc rms}) as a function of integration time as measured in a sensitivity-ordered sub-sample of 47 EBs with a fixed range of baseline length. Upper plot shows the measured image {\sc rms} (blue squares) and the theoretical {\sc rms} expected from the radiometric equation and ALMA sensitivity calculator (red shaded region) \textit{versus} time on source (TOS). The cyan data-point with error-bar shows the median {\sc rms} and the standard deviation of the per-EB continuum map {\sc rms} values. The green solid curve shows the expected {\sc rms} decrease from this median data-point with a simple scaling by the square root of exposure time. Lower plot shows the same information normalized by the green solid curve in the upper plot.}
    \label{fig:sensitivity}
\end{figure}

Since observations were executed in a range of configurations, we have assessed the survey noise properties on a subset of 47 EBs with common baseline ranges (from $70-90$\,m to $8.3$\,km), thus avoiding the contribution from extra baseline coverage (namely at $>8.3\,$km scales). We first determined the noise {\sc rms} on a per-EB basis, and sorted them accordingly. Then the sensitivity assessment was performed on continuum maps having an incremental number of EBs of 1, 2, 4, 8, 16, 32, and 47 EBs (starting from the lowest-noise EB and gradually adding the next EBs in line). We created only ``dirty'' maps (i.e., no deconvolution) using natural weighting (i.e., no image weights applied resulting in highest sensitivity) with the same pixel and image size. Figure~\ref{fig:sensitivity} shows the sensitivity as a function of time on source (TOS), and compares it with the expectation from the radiometric equation \citep[see Equation~9.8 in Section~9.2.1 in][lower region bound]{cortes2025} and the ALMA sensitivity calculator\footnote{\urlasc} (upper region bound). The noise expectation from the latter is slightly higher ($+10\,\%$) than the former. The natural-weighted map combining 47 EBs reaches {\sc rms}=1.7$\,\mu$Jy/beam, while the final continuum map combining all observed 70 EBs reaches {\sc rms}=1.4$\,\mu$Jy/beam. The map created by the standard ALMA pipeline with robust parameter equal to 0.5, reaches 1.9$\,\mu$Jy/beam.

\subsection{Angular scales} \label{ref:scales}

The continuum synthesized beam obtained from combining data over all spectral windows and imaging with Briggs weighting \citep[robust=0.5;][]{Briggs95} is $0.13\arcsec \times 0.15\arcsec$ at PA=33\,deg. At $z_{\rm spec}=1.284$ and 4.699, this implies intrinsic spatial scales of 1.2\,kpc and 0.9\,kpc, respectively. Throughout this article, we also make reference to data cubes and continuum maps imaged with Natural weighting, which results in a larger synthesized beam of $0.22\arcsec \times 0.19\arcsec$ at PA=49\,deg in the continuum map. Natural weighting is preferred since apart from giving higher sensitivity, it is also more desirable for detection by angular scale range.

\begin{figure*}
    \centering
    \includegraphics[width=0.48\textwidth]{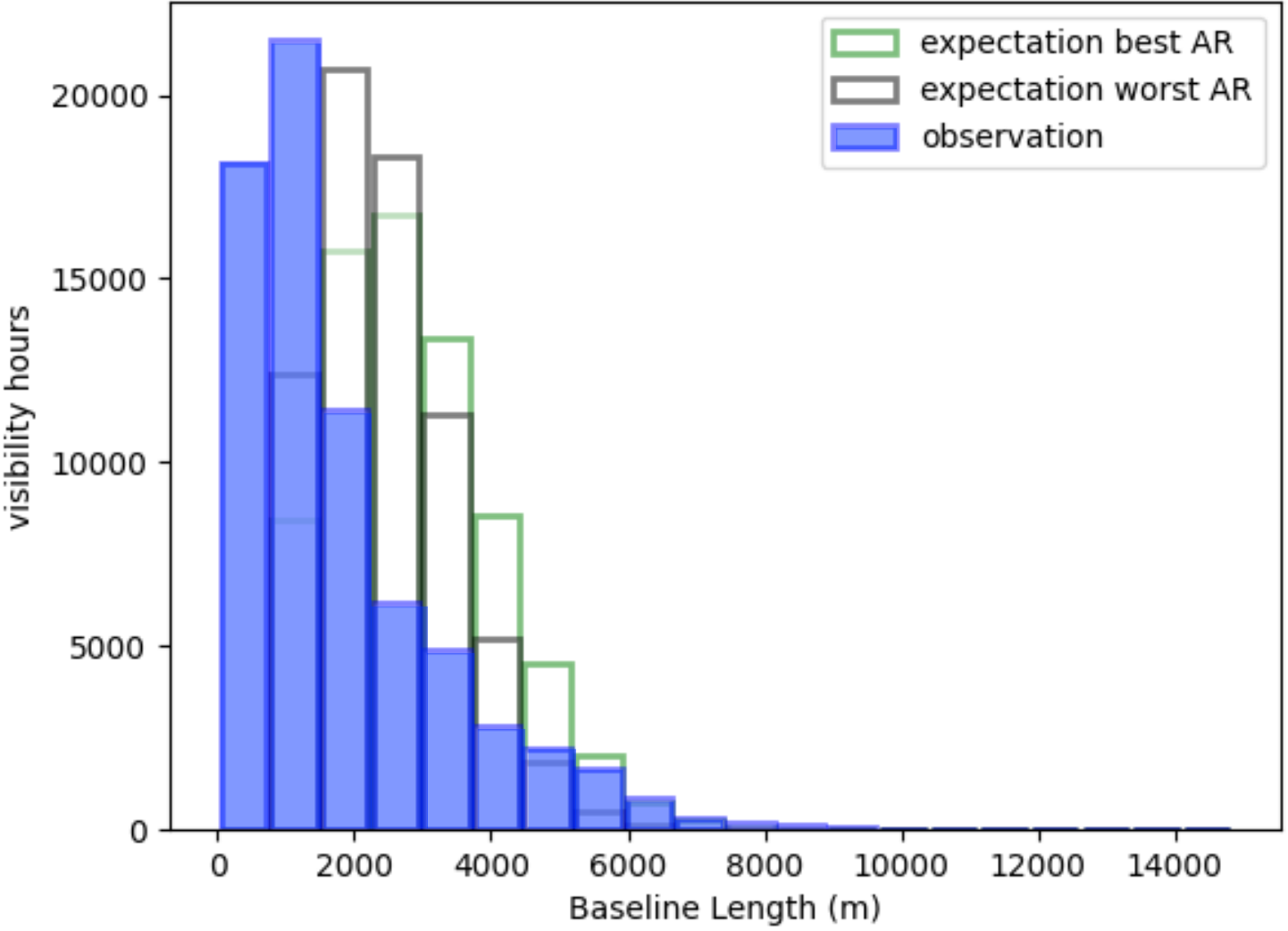}
    \includegraphics[width=0.48\textwidth,trim={0 0 0 0.8cm},clip]{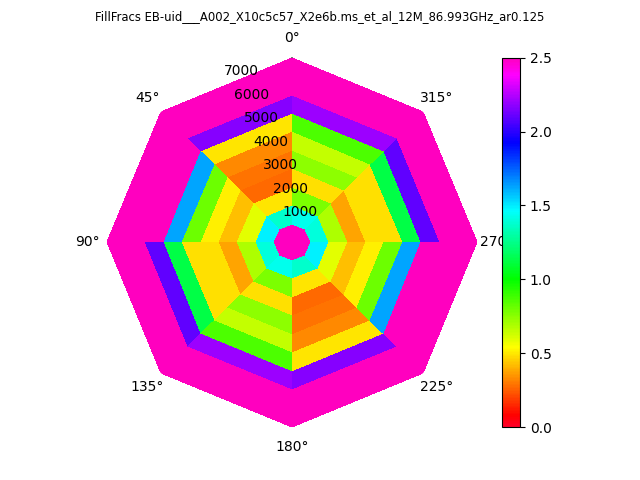}
    \caption{Assessment of the quality of the \textit{uv} coverage of our combined data using the tools developed by \cite{Petry24}. (\textit{Left}) Observed and expected 1-D Baseline Lengths Distributions (BLDs), i.e., histogram of the baseline lengths of the visibilities for the representative frequency channel (87\,GHz) of the dataset (blue) and the corresponding expected histograms for the ideal shape to achieve the most Gaussian PSF given an angular resolution range $\pm20\%$ around the nominal value of $\sim$0.13\arcsec\ (green and grey histograms). (\textit{Right}) Plot of the ratio of observed and expected BLD in 2-D, i.e., also showing baseline orientation. A value of 1.0 indicates that observation and expectation exactly agree. Larger values indicate an over-exposure, smaller values an under-exposure.
    }
    \label{fig:uvcoverage}
\end{figure*}

When covering several configurations without consideration of tailored observing times \citep[see Section~7.8 of ][]{cortes2022}, as in the case of our dataset, the best possible Baseline Length Distribution (BLD) for the most Gaussian beam shape is not achieved. An analysis using the tools developed by \cite{Petry24} shows that the achieved BLD deviates from an ideal shape for the same angular resolution of $\sim0.12\arcsec$ and a maximum recoverable scale of 1\arcsec. In Figure~\ref{fig:uvcoverage} (left-hand side panel), one can see that the observed BLD (blue) has excess sensitivity both at the shortest and the longest baseline lengths and corresponding sensitivity deficits at intermediate baseline lengths between $\sim$1500~m and 5000~m. This is partially a design feature of the ALMA configurations and partially due to the combination of configurations but should have no negative impact on imaging reconstruction. The excess at short baselines is actually beneficial for the detection aspect of the experiment. On the other hand, as already hinted at by the near-circular beam, the right panel of the same figure shows that the azimuthal coverage of our observations is close to perfectly homogeneous (with only very slight under-exposure at azimuth 0--45$^{\circ}$ and baseline lengths $\sim$2000--4000\,m). This is a result of the approximately symmetric array configurations and the wide range of observing hour-angles.

Overall, following the prescription detailed in \cite{Petry24} and using the {\sc assess\_ms~3} software \citep{assess_ms}, we find the following statistics on the final combined MS (scales are referenced to 87\,GHz):
\begin{itemize}
    \item Longest baseline: 14.9\,km
    \item Shortest baseline: 29\,m
    \item Shortest and longest expected BL for a Gaussian beam shape: 152\,m, 7283\,m
    \item 80$^{\rm th}$ percentile baseline length (L$_{80}$) = 2863\,m equivalent to an angular resolution (AR) = 0.14\arcsec
    \item 5$^{\rm th}$ percentile baseline length (L$_{05}$) = 258\,m equivalent to a Maximum Recoverable Scale (MRS) = 2.7\arcsec
    \item a close to homogeneous sensitivity is reached across the entire baseline length range
    \item Baseline orientation is sufficiently homogeneous in all baseline ranges    
\end{itemize}

\section{Results} \label{sec:results}

\subsection{Continuum map}

In order to identify continuum detections at the observed wavelength of 3.2\,mm, we focused on the natural-weighted, primary-beam (PB) uncorrected map with a 0.5\arcsec\ $uv$-taper where the average map noise {\sc rms} is 2.4\,$\mu$Jy/beam and the lowest negative peak in the map is -10.5\,$\mu$Jy/beam. As a result, conservatively, we have only considered as detections those regions in the map with peak fluxes above 10.5\,$\mu$Jy/beam (i.e., $>4.4~\sigma$; resulting in an expected number of false detections in our map of less than one) and within the region where the PB attenuation is $\gtrsim0.5$. In Table~\ref{tab:contdet}, we report the three identified continuum detections and fitting results from the {\it CASA} task {\tt imfit}. Among the three detected sources, only ID-1 is associated with a \textit{Spitzer}/IRAC detection (Appendix~\ref{app:irac}), also showing a faint counterpart in HDFS/\textit{F814W} imaging \citep[$i_{814}=25.5\pm0.03\,$AB;][]{Casertano00}. Moreover, multi-frequency radio observations \citep{Huynh05,Huynh07} with the Australia Telescope Compact Array \citep{Wilson11} show clear detections from $\sim$1\,mJy (SNR=104) at 1.4\,GHz down to $\sim$0.2\,mJy (SNR=19) at 8.7\,GHz. Together with the ALMA detection, we measure a spectral index of $\alpha\sim-0.8$, hence consistent with synchrotron emission. It is not within the scope of this paper to push the continuum detections beyond simple point or compact sources towards extended structures or into the noise limit.

\begin{table*}
\caption{Continuum source detections.}
\label{tab:contdet}      
\centering          
\begin{tabular}{llllllll}
\hline\hline
ID & R.A. (ICRS) & Dec. (ICRS) & PBcorr & $S^{\rm peak}_{\rm 3mm}$ & $S^{\rm integ}_{\rm 3mm}$ & SNR & Size | Orientation \\
& [22:32:ss $\pm$ arcsec] & [-60:mm:ss $\pm$ arcsec] & & [$\mu$Jy/beam] & [$\mu$Jy] & & [arcsec] $\times$ [arcsec] \\
& & & & & & & [degrees]\\
\hline 
1 &  58.631 $\pm$ 0.099 & 33:46.39 $\pm$ 0.13 & 0.806 & 16.9 $\pm$ 2.6 & 36.0 $\pm$ 7.6 & 6.5 & $<$~(2.10 $\times$ 0.32)\\
2 & 56.450 $\pm$ 0.054 & 33:49.398 $\pm$ 0.083 & 0.982 & 13.6 $\pm$ 2.4 & 22.0 $\pm$ 5.9 & 5.6 & (0.74 $\pm$ 0.27) $\times$ (0.30 $\pm$ 0.25) \\
& & & & & & & 14 $\pm$ 28 \\
3 & 59.526 $\pm$ 0.055 & 34:16.886 $\pm$ 0.038 & 0.495 & 28.9 $\pm$ 7.9 & 22.4 $\pm$ 4.4 & 7.9 & (0.93 $\pm$ 0.13) $\times$ (0.75 $\pm$ 0.09) \\
& & & & & & & 85 $\pm$ 22 \\
\hline\hline
\end{tabular}
\tablefoot{The source identification is done in the natural-weighted continuum map with a 0.5'' $uv$-taper. The coordinates, flux, and size estimates were determined with {\it CASA} task {\tt imfit}. Only detections above $4.4~\sigma$ and within the region where PB response is $\gtrsim0.5$ are reported here (see text for details). Reported flux densities are corrected for PB response (cf column PBcorr). Figure~\ref{fig:irac} shows where these sources are located in the field.}
\end{table*}

\subsection{Line emitters} \label{sec:lines}

As mentioned in Section~\ref{sec:specspatcov}, there are at least two galaxy groups identified by MUSE \citep{Bacon15} for which specific CO transitions are covered by the adopted tuning.

\begin{figure}
    \centering
    \includegraphics[width=0.48\textwidth]{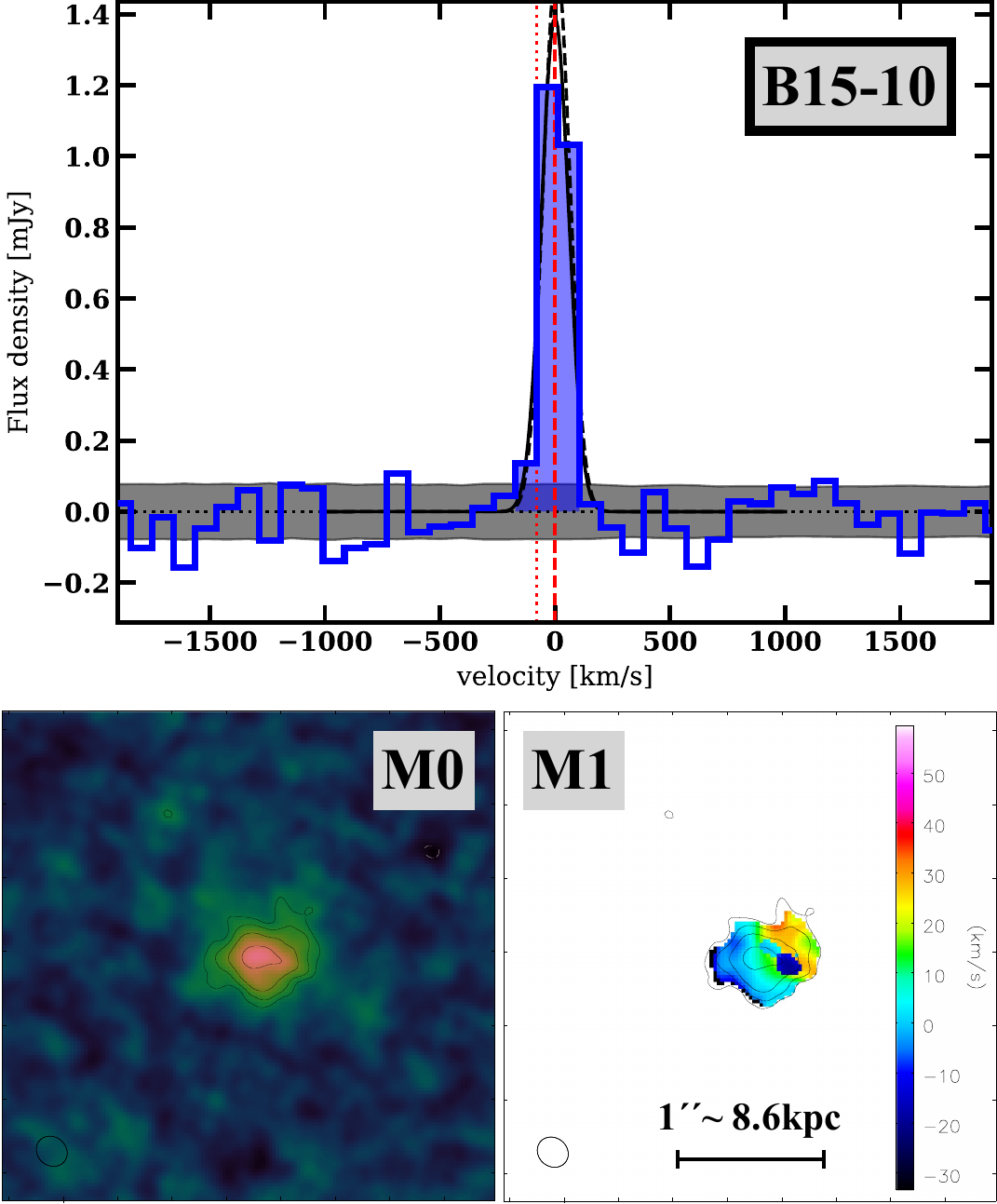}
    \caption{CO\,(2-1) emission towards B15-10. The cube used for the analysis, was imaged with natural weighting and with a smoothing of 4 channels (i.e., resulting in a spectral resolution of $\sim90\,$km/s). We present the spectrum of the detection in the top panel, while the velocity-integrated flux map (moment-0, M0) and the velocity map (moment-1, M1) are shown in the bottom left and right panels, respectively. Both maps have the same size (i.e., 3.4"-wide or $\sim$29\,kpc). The moment-0 contours are overlaid on both maps (black, solid contours indicate levels at $3\times\sqrt(2)^{n}~\times~${\sc rms}, where $n=0,\,1,\,...$, while dashed, white contours indicate $-3\sigma$). The maps are centred at the position reported by \citet{Bacon15}. The top panel with the spectrum displays a filled spectrum within which the flux has been integrated and the moments have been obtained. The red vertical dashed line at zero-velocity assumes the light-weighted centre frequency, while the red dotted vertical line shows the expected frequency assuming the MUSE-derived redshift. The horizontal shaded grey region shows the $\pm1$ times the {\sc std}. A thick line shows the single-component Gaussian best-fit when using the raw spectral-resolution ($\sim24\,$km/s) cube, while a dashed line that when using the 4-channel smoothed cube.}
    \label{fig:b1510}
\end{figure}

\begin{figure}
    \centering
    \includegraphics[width=0.48\textwidth]{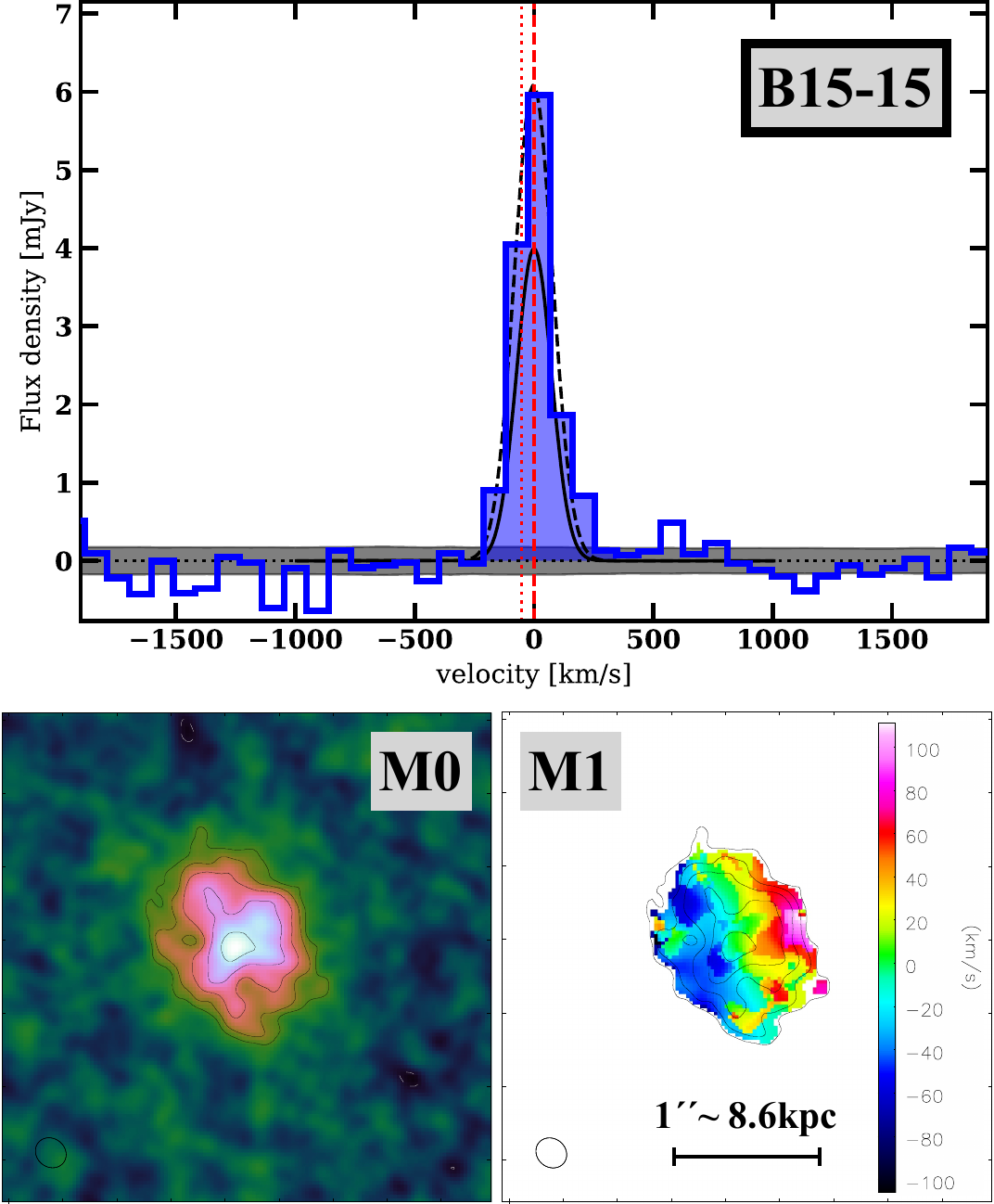}
    \caption{Same as in Figure~\ref{fig:b1510}, but for B15-15.}
    \label{fig:b1515}
\end{figure}

\begin{figure}
    \centering
    \includegraphics[width=0.48\textwidth]{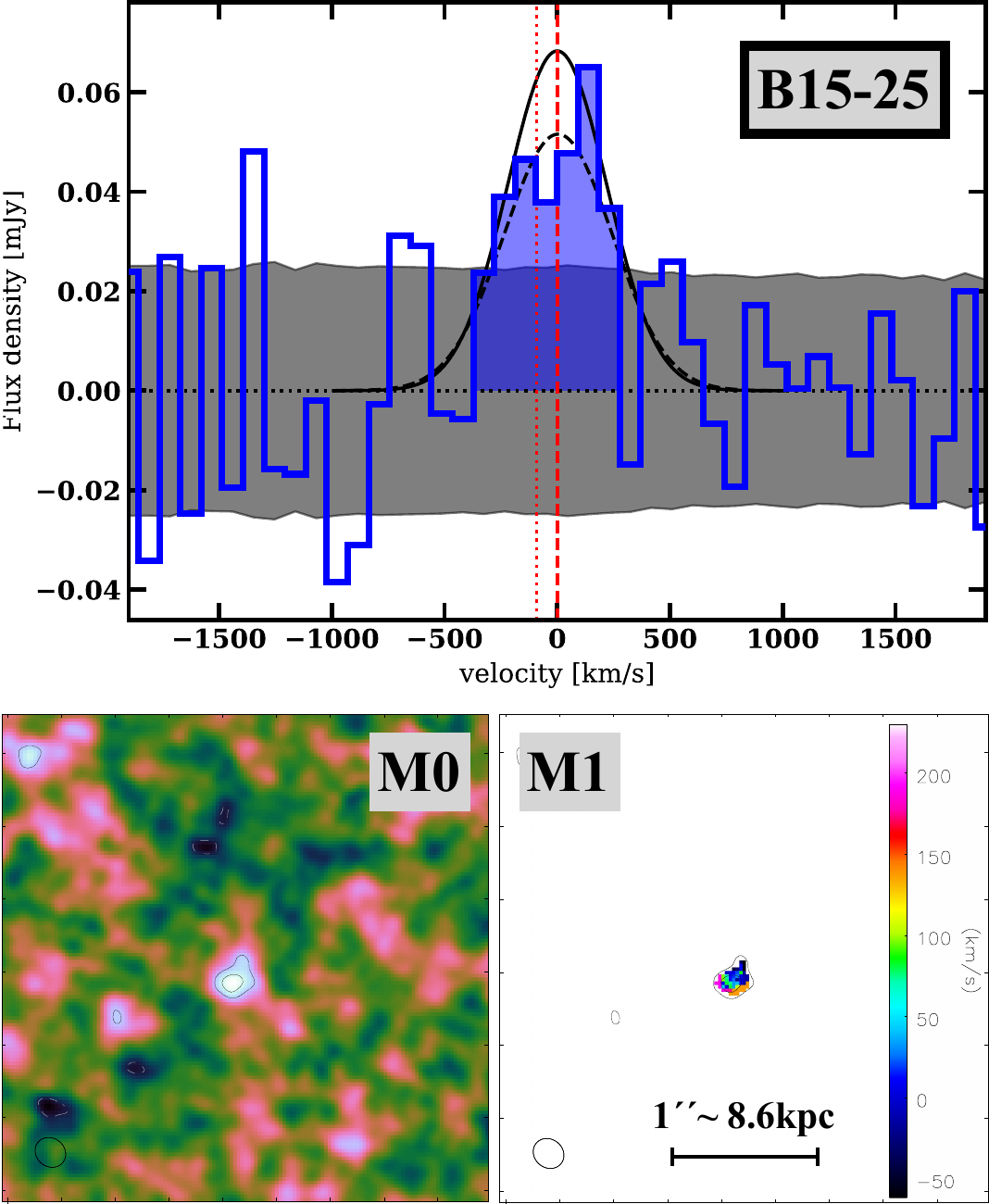}
    \caption{Same as in Figure~\ref{fig:b1510}, but for B15-25.}
    \label{fig:b1525}
\end{figure}

\begin{figure}
    \centering
    \includegraphics[width=0.48\textwidth]{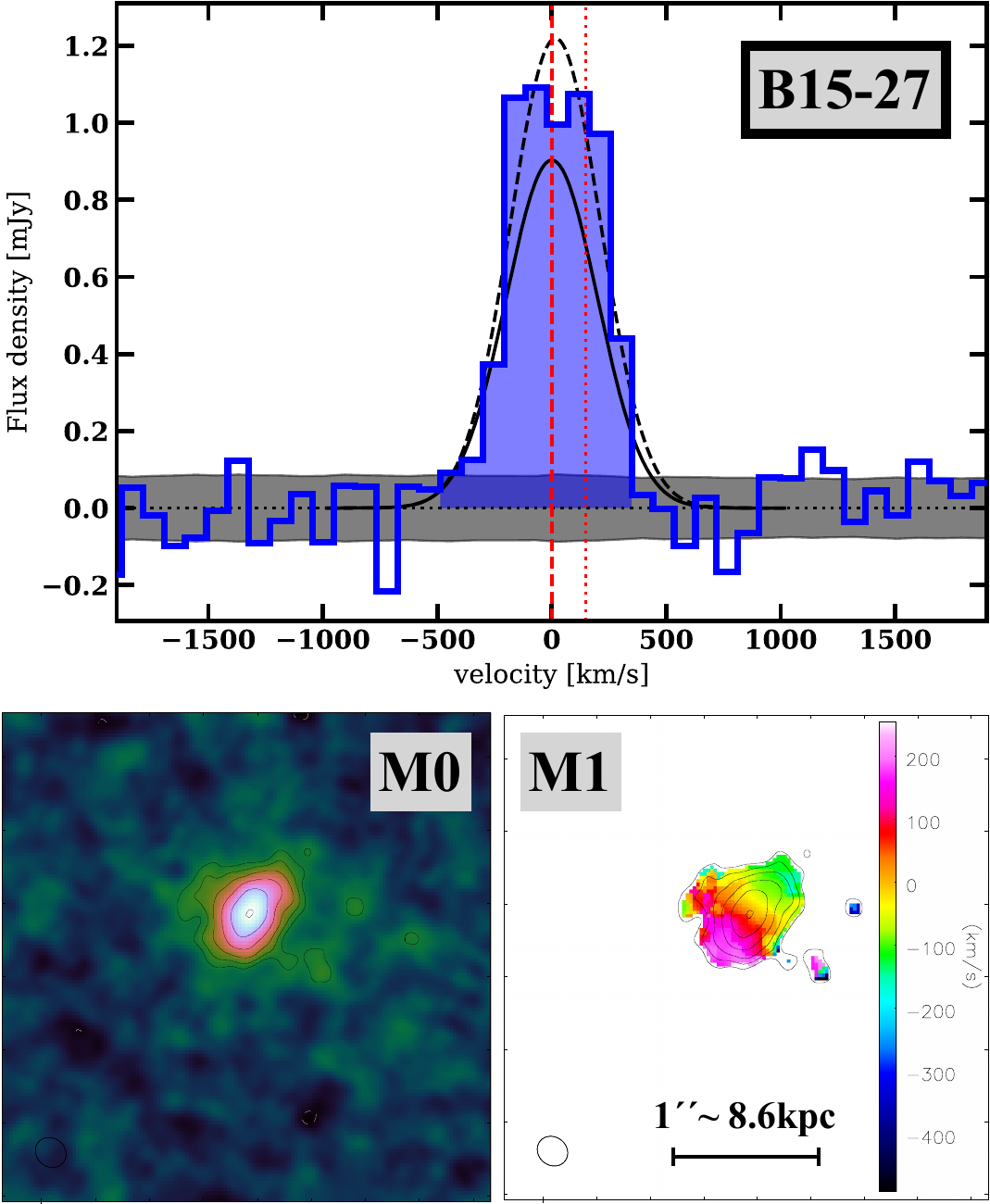}
    \caption{Same as in Figure~\ref{fig:b1510}, but for B15-27.}
    \label{fig:b1527}
\end{figure}

\begin{figure}
    \centering
    \includegraphics[width=0.48\textwidth]{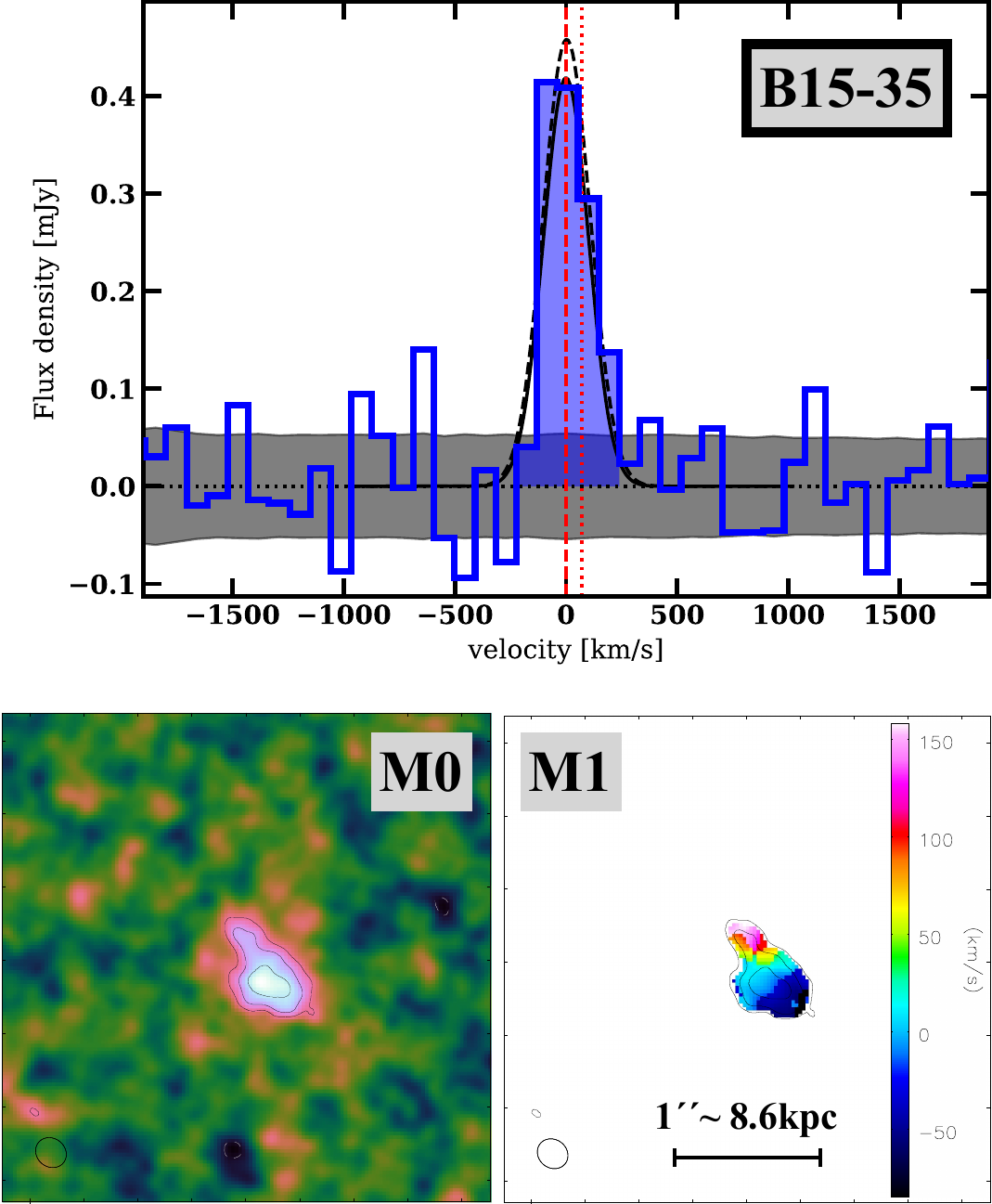}
    \caption{Same as in Figure~\ref{fig:b1510}, but for B15-35.}
    \label{fig:b1535}
\end{figure}

\begin{figure}
    \centering
    \includegraphics[width=0.48\textwidth]{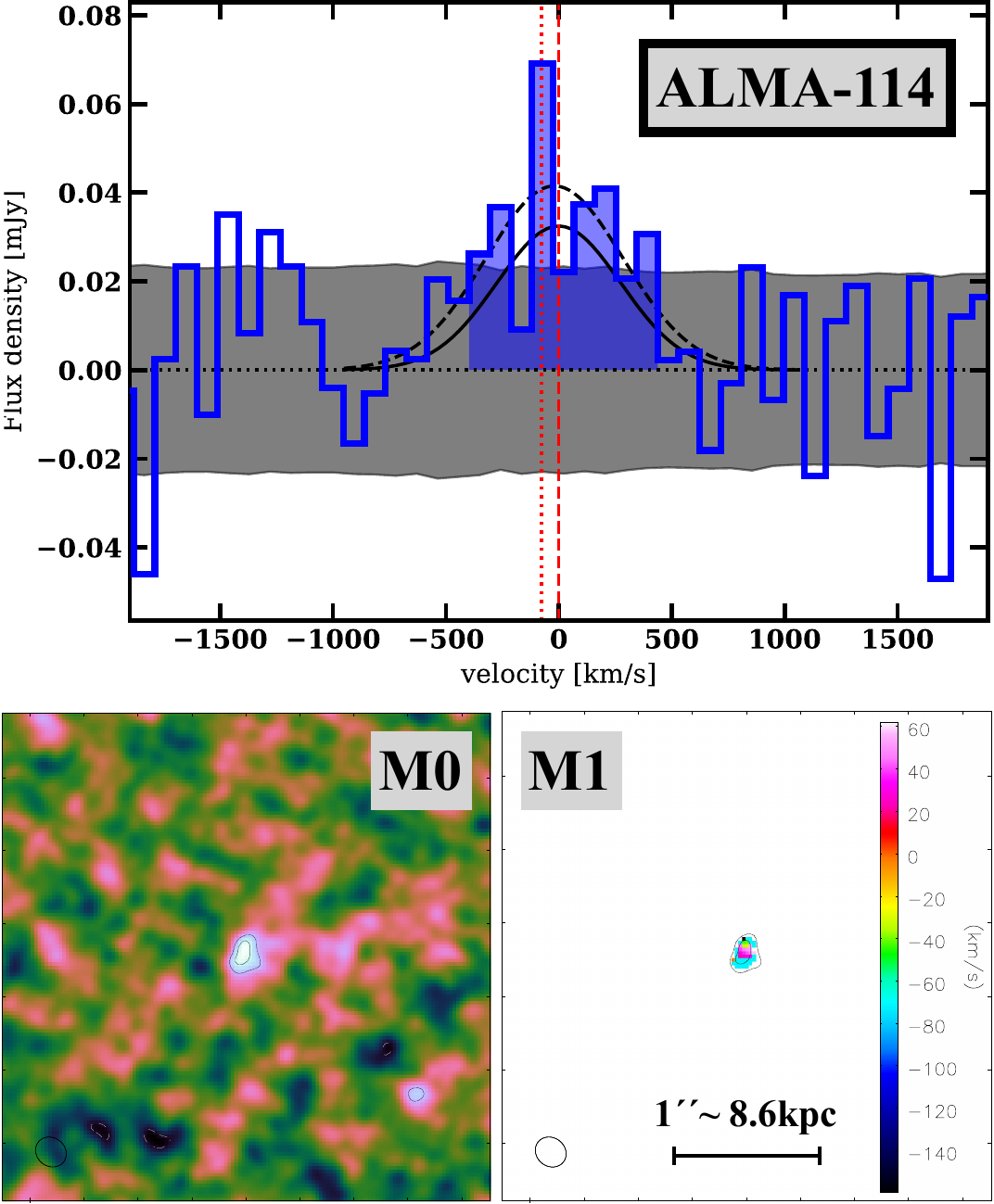}
    \caption{Same as in Figure~\ref{fig:b1510}, but for ALMA-114. We do not detect emission towards the location of B15-114, but we do instead in the North-East towards a source labelled ALMA-114 in this work. The moment-0 map (centred on the new source detection) shows a significant detection (SNR$\sim$6) associated with a broad CO\,(2-1) line detection, very close in redshift to B15-114 (indicated with a vertical dotted line in the spectrum). The lower-SNR feature in the south-west of the image is not associated to the B15-114 source, which is instead located 0.5'' to the south of it.}
    \label{fig:a114}
\end{figure}

With special focus on the highest frequency spectral window, and making use of a natural-weighted cube, we have made a 2-step line search on the velocity-integrated (moment-0) maps towards the location of the MUSE-detected sources at the frequency at which the CO line is expected to be redshifted to based on the redshift reported by MUSE. Briefly, we first started building moment-0 maps within $\pm600\,$km/s around the expected line frequency. This value was done to accommodate potentially broad line profiles as well as significant offsets with respect to the redshift reported by MUSE. This is quite important since the $z=4.7$ group is identified via Lyman-$\alpha$ alone, known to present significant velocity offsets with respect to the host galaxy systemic redshift \citep[e.g.,][]{Verhamme18}. If a significant detection in the moment-0 map was obtained (i.e., above 5 times the median absolute deviation, {\sc mad}), we extracted a preliminary spectrum of the source within the significantly detected region. This spectrum was then used to fine-tune the frequency range within which to retrieve a new moment-0 map. The updated significantly-detected region was subsequently used to extract the final spectrum that was used to determine the spectral range within which to obtain velocity and dispersion maps. The results are displayed in Figures~\ref{fig:b1510} through \ref{fig:b1535}, where we report the CO\,(2-1) line detections towards 5 out of the 9 group members detected by MUSE. Furthermore, we present in Figure~\ref{fig:a114} one serendipitous detection nearby B15-114 which itself does not show CO\,(2-1) emission.

Table~\ref{tab:co21} reports the observed properties of each of the six line detections. The conversion between CO\,(2-1) luminosity ($L^\prime_{\rm CO(2-1)}$) and molecular gas mass (M$_{\rm H_2}$) was done firstly using a factor of $0.75\pm0.11$ to turn $L^\prime_{\rm CO(2-1)}$ to CO\,(1-0) luminosities \citep{Boogaard20} and secondly a CO-to-H$_2$ conversion factor of $\alpha_{\rm CO}=4~{\rm M_\odot / (K.km/s.pc^2)}$ \citep[including Helium;][]{Dunne22}. We acknowledge that the underlying assumption of this value is that these galaxies are metal enriched \citep[$Z/Z_\odot\gtrsim0.5$;][]{Dunne22}. Although there is no metallicity estimates for the members in the targeted group, we can adopt the stellar masses reported by \citet{Contini16} for 4 of the group members (10, 13, 27, and 35 with $\log_{10}({\rm M^*~[M_\odot]})=9.9-10.8$, assuming a \citeauthor{Chabrier03}~\citeyear{Chabrier03} initial mass function) and convert to metallicities assuming the mass-metallicity relation. Either based on observations \citep[adopting local samples, e.g.,][]{Blanc19} or theory \citep[considering redshift evolution, e.g.,][]{Ma16} we obtain $\frac{Z_{gas}}{Z_\odot}=0.5-0.7$ or 0.3--0.6, respectively. This shows that the sample is mostly still within the applicable metallicity range.

We finally note that half of the detected sample shows $M_{H_2}$ values at and below the H$_2$ mass limit reached by the deepest ALMA line surveys at similar redshifts so far \citep[M$_{\rm H_2}=0.63-9.7\times10^{10}$\,M$_\odot$ at $1.09<z<1.55$;][]{Decarli19}.

\begin{table*}
\caption{CO\,(2-1) line detections towards the galaxy group at $z_{\rm spec}=1.284$}             
\label{tab:co21}      
\centering          
\begin{tabular}{ccccrrrrr} 
\hline\hline
Source\tablefootmark{a} & HDFS\tablefootmark{b} & RA$_{\rm J2000}$ & Dec.$_{\rm ~J2000}$ & redshift & $\nu_{\rm obs}$ & $\Delta v~S_\nu$ & $L^\prime_{\rm CO(2-1)}$ & M$_{\rm H_2}$ \\
& & [h:m:s] & [d:m:s] & & [GHz] & [mJy.km/s] & [10$^9$~K.km/s.pc$^2$] & [10$^9$~M$_\odot$] \\
\hline 
B15-10 & 1539 & 22:32:53.018 & -60:33:28.55 & 1.2846 & 100.9086 & 219$\pm$14 & 5.02$\pm$0.31 & 15.1$\pm$2.4 \\
B15-15 & 1334 & 22:32:52.883 & -60:33:17.00 & 1.2842 & 100.9259 & 1275$\pm$69 & 29.2$\pm$1.6 & 88$\pm$14 \\
B15-25 & 1557 & 22:32:54.991 & -60:33:29.08 & 1.2834 & 100.9641 & 27.5$\pm$4.3 & 0.628$\pm$0.098 & 1.89$\pm$0.40 \\
B15-27 & 1650 & 22:32:53.407 & -60:33:33.06 & 1.2843 & 100.9228 & 579$\pm$33 & 13.24$\pm$0.75 & 39.7$\pm$6.2 \\
B15-35 & 1800 & 22:32:53.979 & -60:33:41.66 & 1.2801 & 101.1074 & 120.0$\pm$9.5 & 2.73$\pm$0.22 & 8.2$\pm$1.4 \\
ALMA-114 & --- & 22:32:54.403 & -60:33:45.70 & 1.2834 & 100.9618 & 30.5$\pm$5.0 & 0.70$\pm$0.12 & 2.09$\pm$0.46 \\
\hline
\end{tabular}
\tablefoot{
\tablefoottext{a}{IDs in \citet{Bacon15}.}
\tablefoottext{b}{IDs in \citet{Casertano00}.}
}
\end{table*}

We have also fitted the detected lines with single-component Gaussian profiles. The best-fit solutions are shown in the spectra in the top panels in Figures\,\ref{fig:b1510} through \ref{fig:a114}. The best-fit is shown by a thick line for the raw spectral-resolution (~24 km/s) cube and a dashed line for the 4-channel smoothed cube. Table\,\ref{tab:linefits} reports the fitting results, while Table\,\ref{tab:zcomp} reveals the redshift determination differences between CO and optical estimates.

\begin{table*}
\caption{Gaussian fit results to the line detections}             
\label{tab:linefits}      
\centering          
\begin{tabular}{c|rrr|rrr}  
\hline\hline
 & \multicolumn{3}{c|}{Smoothed Spectrum} & \multicolumn{3}{c}{Raw Spectrum} \\
 & \multicolumn{3}{c|}{[$\Delta v\sim90\,{\rm km/s}$]} & \multicolumn{3}{c}{[$\Delta v\sim24\,{\rm km/s}$]} \\
\hline
Source & $S_\nu^{\rm peak}$ & FWHM & $\Delta v~S_\nu$ & $S_\nu^{\rm peak}$ & FWHM & $\Delta v~S_\nu$ \\

& [mJy] & [km/s] & [Jy.km/s] & [mJy] & [km/s] & [Jy.km/s] \\
\hline 
B15-10 & 1.54$\pm$0.14 & 96$\pm$12 & 0.158$\pm$0.024 & 1.39$\pm$0.13 & 98$\pm$10 & 0.146$\pm$0.020 \\
B15-15 & 6.08$\pm$0.18 & 132.7$\pm$4.4 & 0.860$\pm$0.038 & 4.00$\pm$0.26 & 121.5$\pm$9.1 & 0.517$\pm$0.051 \\
B15-25 & 0.052$\pm$0.014 & 400$\pm$130 & 0.0220$\pm$0.0092 & 0.068$\pm$0.019 & 370$\pm$120 & 0.027$\pm$0.011 \\
B15-27 & 1.220$\pm$0.054 & 328$\pm$16 & 0.426$\pm$0.028 & 0.903$\pm$0.076 & 331$\pm$32 & 0.318$\pm$0.041 \\
B15-35 & 0.458$\pm$0.046 & 177$\pm$20 & 0.086$\pm$0.013 & 0.419$\pm$0.067 & 169$\pm$31 & 0.075$\pm$0.018 \\
ALMA-114 & 0.042$\pm$0.012 & 510$\pm$170 & 0.0224$\pm$0.0097 & 0.032$\pm$0.014 & 460$\pm$230 & 0.016$\pm$0.010 \\
\hline
\end{tabular}
\tablefoot{The fits resulting from adopting the raw and smoothed spectra are displayed in Figures~\ref{fig:b1510} through \ref{fig:a114} as solid and dashed black lines, respectively.}
\end{table*}

\begin{table*}
\caption{Redshift comparison between ALMA and MUSE}             
\label{tab:zcomp}      
\centering          
\begin{tabular}{crrrcr}
\hline\hline
Source & Freq$_{\rm cent}$\,[GHz] & $z_{\rm CO}$ & $z_{\rm opt}$ & lines$_{\rm opt}$ & $\Delta\,v_{\rm opt}$\,[km/s] \\
\hline 
B15-10 & 100.9093$\pm$0.0021 & 1.284607$\pm$2.7E-5 & 1.2840 & [OII], MgII & -80 \\
B15-15 & 100.9275$\pm$0.0019 & 1.284195$\pm$2.4E-5 & 1.2838 & [OII], MgII & -52 \\
B15-25 & 100.967$\pm$0.024 & 1.28330$\pm$3.0E-4 & 1.2826 & [OII],[NeIII] & -92 \\
B15-27 & 100.9285$\pm$0.0065 & 1.284171$\pm$8.2E-5 & 1.2853 & [OII], MgII & +148 \\
B15-35 & 101.1107$\pm$0.0063 & 1.280056$\pm$8.0E-5 & 1.2806 & [OII], MgII & +72 \\
ALMA-114 & 100.959$\pm$0.046 & 1.28349$\pm$5.9E-4 & --- & ---  & ---\\
\hline
\end{tabular}
\tablefoot{
The Freq$_{\rm cent}$ and $z_{\rm CO}$ columns refer to the best-fit line centroid found in the raw spectral resolution cube (thick line Gaussian fit in the bottom-left panels in Figures\,\ref{fig:b1510} through \ref{fig:a114}). The $z_{\rm opt}$ and lines$_{\rm opt}$ columns are extract from \cite{Bacon15}.
}
\end{table*}

Overall, the diversity found in this group is clear as well as the proximity of the galaxies to each other. For reference, the brightest three galaxies we find (B15-10, 15, 27) are within $\sim$50 to $\sim$140\,kpc (projected) from each other, and \citet{Contini16} finds evidence of low-surface brightness emission between B15-10 and B15-27 hinting that these two galaxies may have already interacted. B15-27 is an interesting case on its own when comparing its dynamical map with that based on [OII]$_{3729}$ \citep{Contini16} and the rest-frame UV map (Appendix~\ref{app:hst}). While this galaxy shows a very clean rotation pattern, the CO(2-1) velocity-integrated flux centroid is clearly offset from its rest-frame UV counterparts. The latter actually seem to show a very disturbed system, with the [OII]$_{3729}$ velocity map showing a clear extension to the West. This is in line with the location of one or two close-by companion clumps detected in CO(2-1) ($\sim5\,$kpc away and offset in velocity up to $-500\,$km/s; one of the clumps potentially showing a rotation pattern too). We also find hints for a higher-velocity feature (at 100--150\,km/s) to the North-East of B15-35, but, despite the clear alignment between the velocity patterns between ionized and molecular gas, \citet{Contini16} does not report significant evidences for such a feature in the [OII]$_{3729}$ velocity map. Given the multi-wavelength richness towards the detected sources, we differ a more complete dynamical analysis to a future work.

Moreover, we find a significant ($\sigma\sim6$) serendipitous CO\,(2-1) detection to the North-East of B15-114 (which we label ALMA-114). B15-114 itself remains undetected in CO\,(2-1), but its MUSE --- [OII]-derived --- redshift of $z_{\rm spec}=1.2844$ is very close to the one obtained for this companion, $z_{\rm spec}=1.28349\pm0.00059$.

Finally, we have not detected any line emission towards the other groups listed in Table~\ref{tab:grpcov} or within the whole field. It is beyond the scope of this manuscript to pursue more detailed detection analysis (e.g., stacking) especially towards the high-redshift group for which the systemic redshifts are uncertain. We do note that the shallower, yet wider, ASPECS-3mm survey \citep{Decarli19} also does not report any CO(5-4) detections at $z>4$. The COLDz survey \citep{Riechers19} does report detections of CO(2-1) at $z\sim5$ towards three previously known dusty star-forming galaxies (DSFGs). The CO(5-4) line fluxes reported therein for the same galaxies would have been detectable in our survey. This also shows that this MUSE-identified galaxy group does not include such DSFGs within $±200\,$kpc (projected) from the identified group members. More reliable conclusions on source properties for this group would require the knowledge of the systemic redshift, which Lyman-$\alpha$ line emission does not guarantee.

We have also made use of the Source Finding Application \citep[version 2.5.1, SoFiA-2;][]{Serra15,Westmeier21}\footnote{\urlsof} to blindly search for emission lines in each of the four spectral windows. However, only four of the detections reported before have been automatically considered significant by SoFiA-2, showing that in this case, it is better to know where the sources are located with an initial guess of the systemic redshift.

\subsection{Cosmic Molecular gas mass density in group environments} \label{sec:massdens}

Despite the small projected survey size (1\,arcmin$^2$), we attempt to estimate the cosmic molecular gas density ($\rho_{\rm H_2}$) from the galaxy population in group environments. We use this [OII]$_{3727,3729}$-identified group at $z_{\rm spec}=1.284$ as reference, hence assumed to be representative.

We adopt the $1/V_{\rm max}$ formalism \cite{Schmidt68}. The $z_{\rm min}$ and $z_{\rm max}$ were estimated with two slightly different approaches. What is common between both limits is the consideration of the MUSE sensitivity profile (or pass band) from 480 to 930\,nm that allows us to take into account the sensitivity dips \citep[that translate to reduced completeness;][]{Drake17} in the whole spectral range. We obtained it from the VLT/MUSE Exposure Time Calculator, emulating the observing conditions of the MUSE observations taken around 29$^{\rm th}$ July 2014. The obtained trend was smoothed to the line FWHM reported in Table~\ref{tab:linefits}.

One should be reminded that the redshift estimate of a source for which only the [OII] doublet is detected depends on resolving the doublet (which MUSE is capable of) and detecting the fainter line at rest-frame 3727\,\AA. So we adopt it as reference and a 3729 to 3727\,\AA\ line ratio of 1.3 \citep{Kaasinen17}, since only the combined flux of the doublet is being reported in \cite{Bacon15}.

The $z_{\rm min}$ was estimated only based on the MUSE coverage of [OII]$_{3727,3729}$ together with the sensitivity profile guaranteeing a $SNR>5$, resulting in $z_{\rm min}=0.289$.

On the other hand, $z_{\rm max}$ was estimated taking into consideration both the [OII]$_{3727,3729}$ and CO\,(2-1) integrated line fluxes. Briefly, $z_{\rm max}$ is limited to MUSE coverage of [OII]$_{3727,3729}$ at $z\sim1.49$. On a case-by-case basis, this value may be of course lower depending on the observed [OII]$_{3727,3729}$ and CO\,(2-1) properties.

There are two additional corrections we have adopted:
\begin{itemize}
    \item One relates to the strong sensitivity dips in MUSE sensitivity profile (e.g, at $\sim860-870\,$nm), which for a narrow redshift range may result in a non-detection \citep[hence lower completeness;][]{Drake17}, but at higher redshifts it is considered detectable ``again''. To account for this we corrected the full initial volume (from $z_{\rm min}$ to $z_{\rm max}$) by the fraction of bins in that range that would yield a detection. This value is found to be between 0.924 and 0.975, and hence a small correction in this case.
    \item Also, the group identification was done for groups of 3 or more members, so this group would be identified as such until the ``last'' three members would be detectable. In this case, these would be B15-10, B15-15, and B15-35, and $z_{\rm max}$ is caped to $z=1.4935$.
\end{itemize}

We do not attempt to correct for intrinsic evolution of this group, in other words, for the fact that the star formation rate density (hence [OII] emission related to star formation) strongly declines from $z=1.49$ to $z=0.289$, hence potentially our ability to identify these sort of numerous groups. But we do note that, in doing so, one would increase further the reported molecular gas mass volumetric density.

Finally, we must consider cosmic variance and low-number statistics. To account for this we followed \citet{TrentiStiavelli08}, by using their calculator\footnotetext{Version 1.03 (24 July 2020): \urlcal} to compute the total fractional error (Poisson uncertainty and cosmic variance) on number counts based on redshift range, field size, and number of sources. We find that this value is about 0.5 in our case. Given the low-number statistics, whenever we refer to statistical uncertainty, it is estimated as $\pm0.5+\sqrt{N+0.25}$\footnote{This alternative is proposed by the Collider Detector at Fermilab Statistics Committee and also accessible within the {\sc astropy.stats} package, in the function {\sc poisson\_conf\_interval}, by setting  {\sc interval=`pearson'}.}, where the negative and positive signs refer to the lower and upper errors, respectively.

Nevertheless, despite what is mentioned in the previous two paragraphs, we must highlight that other surveys conducted with MUSE down to similar or deeper levels \citep{Bacon17,Bacon23,Fossati19} also show groups at similar redshifts comprising numerous members. For instance, \citet{Fossati19} reports two groups with more than 10 members at $z_{\rm spec}\sim0.678$ and 1.053. The catalogues provided by \citet{Bacon23} based on the deepest MUSE survey ever taken also show two clear redshift peaks at $z_{\rm spec}\sim0.659$ and 1.089. Also in \citet{Bacon15}, apart from the group of focus in this manuscript, there is another one comprising 7 members at $z_{\rm spec}\sim0.564$. This thus seems to show that [OII]-identified groups may be a common observable when the Universe had an age between 5.5 and 7.5\,Gyr.

\begin{figure}
    \centering
    \includegraphics[width=\linewidth]{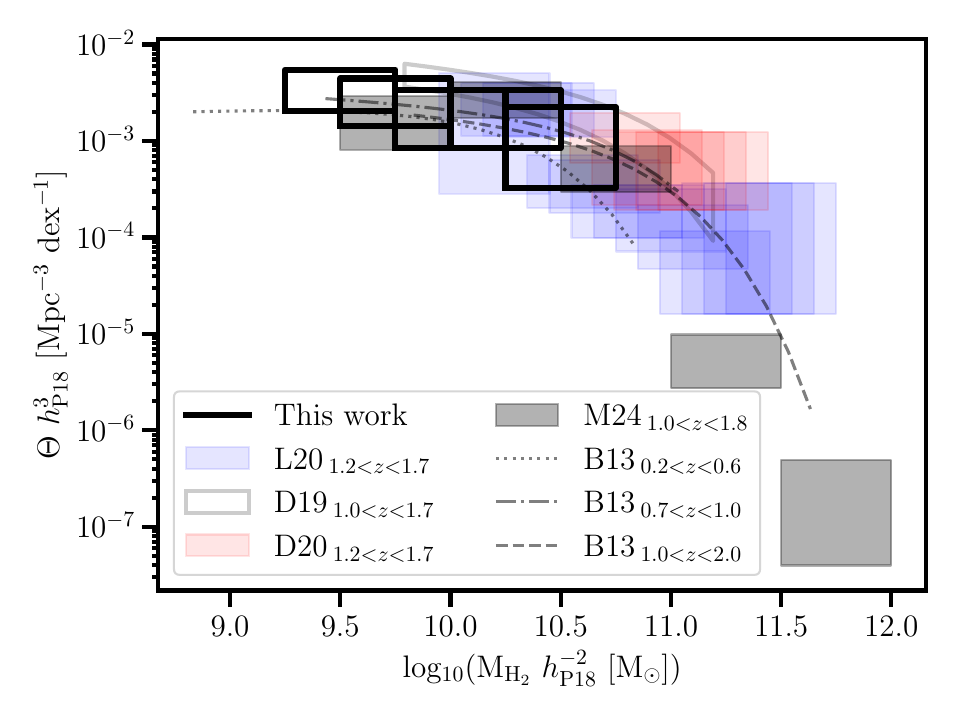}
    \caption{The molecular gas mass function derived from the observed group at $z_{\rm spec}=1.284$ compared to works in the literature by \citet[][D19 and D20, at $1.0<z<1.7$]{Decarli19,Decarli20}, \citet[][L20, at $1.2<z<1.7$]{Lenkic20}, and \citet[][M24, at $1.0<z<1.8$]{Messias24}. The results by \citet[][different line types for different redshift ranges]{Berta13} are derived from empirical scaling relations, instead of inferring M$_{\rm H_2}$ from CO as the previously cited works. The empty boxes show the results from our study. These are 0.5\,dex wide in molecular gas mass (centred at each step of 0.25\,dex), while the vertical width shows the statistical uncertainty.}
    \label{fig:massfunc}
\end{figure}

The molecular Hydrogen mass function (MF) obtained for this group at $z_{spec}=1.284$ is presented in Figure~\ref{fig:massfunc}. We compare this result with literature work representative of the galaxy population in the field environment by \citet[][$1.0<z<1.7$]{Decarli19,Decarli20}, \citet[][$1.2<z<1.7$]{Lenkic20}, and \citet[][$1.0<z<1.8$]{Messias24}. We additionally show results reported by \citet{Berta13} who adopted two different scalings with star formation rate to derive the H$_2$ MF at different redshift ranges in-between $0.2<z<2.0$. All data have been made consistent by adopting the same CO-to-H$_2$ conversion factor \citep[$\alpha_{\rm CO}=4$ including Helium fraction;][]{Dunne22} and cosmology. The figure shows that group and field environments seem to show similar MFs (within the current uncertainties), but, more noticeably, our survey is going deeper almost by 0.5\,dex in molecular mass at these redshifts than ASPECS \citep{Decarli19,Decarli20} and \citet{Berta13}. At these levels, the results show that the molecular gas MF is still rising with decreasing luminosity, yet still consistent with the flat light-end of the H$_2$ MF observed as late as $0.2<z<0.6$.

\begin{figure}
    \centering
    \includegraphics[width=\linewidth]{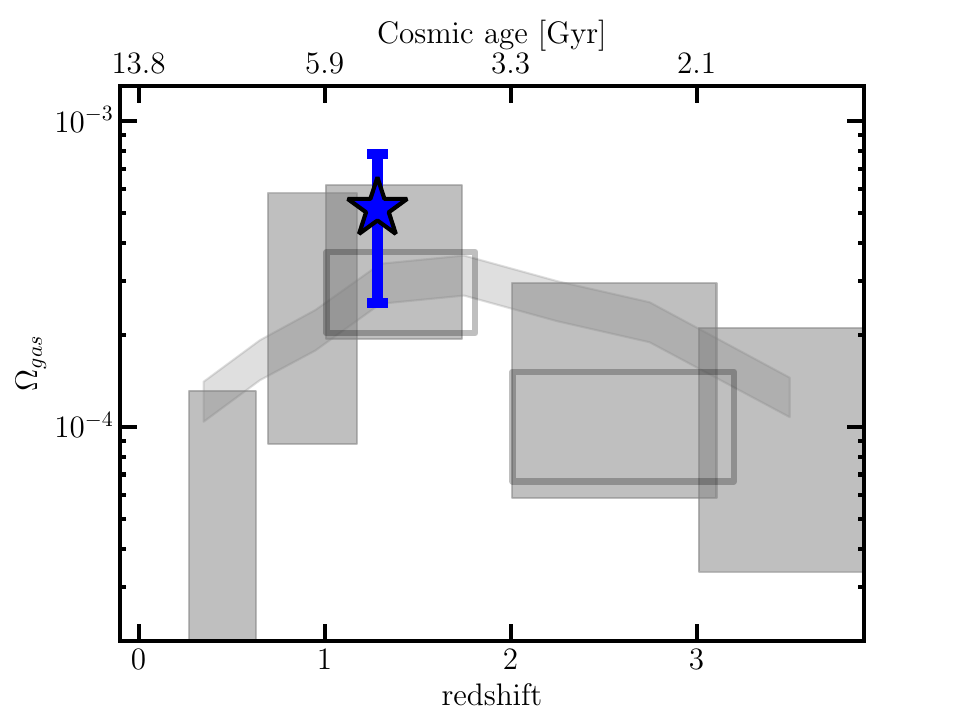}
    \caption{The cosmic molecular gas mass density in group environments derived making use of group identified by MUSE and detected by ALMA in CO\,(2-1) at $z=1.284$ (blue data point and error-bar). For reference, literature estimates are also included. The CO-derived estimates reported by \citet{Decarli20} from $z=0.271$ to 4.475 appear as grey filled boxes, while those at $1\lesssim z \lesssim 3$ reported by \citet{Messias24} appear as empty boxes. The sub-mm continuum derived trend from  \citet{Scoville17} is displayed as a continuous grey filled region. The cosmic mass density is plotted as a ratio to the cosmic critical density.}
    \label{fig:massdens}
\end{figure}

In Figure~\ref{fig:massdens} we report the cosmic molecular gas mass density in group environments compared to results from the literature either based on surveys directly detecting CO rotational transitions \citep[][grey-line boxes]{Decarli20,Messias24} or sub-mm continuum relations \citep[][continuous grey-line region]{Scoville17}. What must be highlighted here is that the molecular gas mass densities recovered by this survey\footnote{We note that removing ALMA-114 from the analysis would only reduce $\rho_{\rm H_2}$ by 2\%.} are comparable to those of blind CO surveys in the literature. The implications for this are discussed in Section~\ref{sec:conc}.

\section{Conclusions} \label{sec:conc}

In this article we describe a $1\times1\,$arcmin$^2$ deep single-tuning survey at sub-arcsec scales conducted by ALMA in the \textit{Hubble} Deep Field South (HDFS). The spectral tuning was especially chosen to trace CO transitions in galaxies belonging to two groups identified by MUSE at $z_{\rm spec}=1.284$ and 4.699 (Section~\ref{sec:obs}).

This survey was conducted as an ALMA Observatory Project (filler in nature) that resulted in a total time on source of 61\,h, yielding a continuum sensitivity of {\sc rms}=1.4\,$\mu$Jy/beam (adopting natural weighting; Section~\ref{sec:depth}). The unprecedented $uv$-coverage range for a mm survey yields a spatial resolution of $0.13\arcsec \times 0.15\arcsec$ (adopting Briggs weighting with robust=0.5) and maximum recoverable scales of 1--2\,arcsec (Section~\ref{ref:scales}).

We report CO\,(2-1) detections toward 5 out of the 9 members of the galaxy group at $z_{\rm spec}=1.284$, and further identified a previously unknown member of the same group which shows no optical or NIR counterpart (Section~\ref{sec:lines}). Half of the sample detected by ALMA have expected molecular gas masses at and below the mass-cut reached by previous deep spectral-scan surveys covering similar redshifts. It is worth noting that the gas detections display a wide range of diversity with no two sources alike in terms of their size and dynamical states. Only one shows a smooth rotation pattern, but also shows evidence for two satellite features and it shows a clear offset with respect to its rest-frame UV-optical emission; the observed line widths are varied; and the sizes vary from point-like to 10\,kpc.

Based on the CO\,(2-1) detections toward this group, we attempted an estimate of the molecular gas mass function (MF) and the cosmic molecular gas mass density ($\Omega_{\rm H2}$) in group environments. We find evidence that group environments have MFs undistinguishable from those of field galaxies (within the observed uncertainties), and, to the survey depth, the MFs show no hint of a decrease in number density with decreasing molecular gas, but are still consistent with the flat light-end of the MFs observed at $z<1$. We find that the recovered $\Omega_{\rm H2}$ is comparable to that of populations with larger H$_2$ contents found in wider and shallower surveys. Given the small projected survey size we acknowledge that there is room for a large uncertainty in the result, however, we do note that such group seems not to be a rare observable at similar redshifts in other deep MUSE fields. Hence, this gives us confidence that our result may indeed be representative of group environments when the Universe was approximately half its current age (Section\,\ref{sec:massdens}). Moreover, this should be evidence that our current knowledge of $\Omega_{\rm H2}$ and its evolution back in time is still rather incomplete at $z>1$. In other words, we still do not know when $\Omega_{\rm H2}$ peaked, and any estimate of how much it has decreased since $z\sim1-2$ needs to be taken as a lower limit. Nevertheless, the potential no-evolution scenario of the H$_2$ MF light-end (M$_{\rm H2}\lesssim10^{10}\,$M$_\odot$) between $z\sim0.2$ and $\sim$1.3 may help compensating for the increased incompleteness affecting current surveys at earlier cosmic times.

Finally, we detect three significant sources in the continuum map (SNR), where only one shows multi-wavelength counterparts from rest-frame UV to radio frequencies. The other two continuum detections are expected to be thermal in nature, since a synchrotron SED should result in clear detections in existing data at 1.4\,GHz.

\begin{acknowledgements}
This paper makes use of the following ALMA data: ADS/JAO.ALMA\#2022.A.00034.S. ALMA is a partnership of ESO (representing its member states), NSF (USA) and NINS (Japan), together with NRC (Canada), NSTC and ASIAA (Taiwan), and KASI (Republic of Korea), in cooperation with the Republic of Chile. The Joint ALMA Observatory is operated by ESO, AUI/NRAO and NAOJ.

The National Radio Astronomy Observatory is a facility of the National Science Foundation operated under a cooperative agreement by Associated Universities, Inc.

This research made use of {\sc ipython} \citep{PerezGranger07}, {\sc numpy} \citep{Walt11}, {\sc matplotlib} \citep{Hunter07}, {\sc scipy} \citep{Virtanen20}, {\sc astropy} \citep[a community-developed core {\sc python} package for Astronomy,][]{Astropy13}, {\sc topcat} \citep{Taylor05}, APLpy \citep[an open-source plotting package for Python,][]{Robitaille12}.

The team appreciates the constructive feedback provided by the anonymous referee that pushed us toward a significantly improved version of the manuscript.

\end{acknowledgements}

\bibliographystyle{aa}
\bibliography{aa52831-24.bib}

@ARTICLE{assess_ms,
       author = {{Petry}, D. and {Diaz Trigo}, M. and {Kneissl}, R. and {Toledo}, I. and {Facchini}, S. and {Lorber}, D. and {Bonanomi}, F. and {Hacar}, A.},
        title = "{Assess\_ms 3 - the ALMA uv Coverage Assessment Tool}",
         year = 2025,
        month = aug,
          doi = {10.5281/zenodo.16682282},
       url = {https://zenodo.org/records/16682282}
}

@ARTICLE{Blanc19,
       author = {{Blanc}, Guillermo A. and {Lu}, Yu and {Benson}, Andrew and {Katsianis}, Antonios and {Barraza}, Marcelo},
        title = "{A Characteristic Mass Scale in the Mass-Metallicity Relation of Galaxies}",
      journal = {\apj},
         year = 2019,
        month = may,
       volume = {877},
       number = {1},
          eid = {6},
        pages = {6},
          doi = {10.3847/1538-4357/ab16ec},
archivePrefix = {arXiv},
       eprint = {1904.02721},
 primaryClass = {astro-ph.GA},
       adsurl = {https://ui.adsabs.harvard.edu/abs/2019ApJ...877....6B}
}

@ARTICLE{Ma16,
       author = {{Ma}, Xiangcheng and {Hopkins}, Philip F. and {Faucher-Gigu{\`e}re}, Claude-Andr{\'e} and {Zolman}, Nick and {Muratov}, Alexander L. and {Kere{\v{s}}}, Du{\v{s}}an and {Quataert}, Eliot},
        title = "{The origin and evolution of the galaxy mass-metallicity relation}",
      journal = {\mnras},
         year = 2016,
        month = feb,
       volume = {456},
       number = {2},
        pages = {2140-2156},
          doi = {10.1093/mnras/stv2659},
archivePrefix = {arXiv},
       eprint = {1504.02097},
 primaryClass = {astro-ph.GA},
       adsurl = {https://ui.adsabs.harvard.edu/abs/2016MNRAS.456.2140M}
}

@ARTICLE{Chabrier03,
       author = {{Chabrier}, Gilles},
        title = "{Galactic Stellar and Substellar Initial Mass Function}",
      journal = {\pasp},
         year = 2003,
        month = jul,
       volume = {115},
       number = {809},
        pages = {763-795},
          doi = {10.1086/376392},
archivePrefix = {arXiv},
       eprint = {astro-ph/0304382},
 primaryClass = {astro-ph},
       adsurl = {https://ui.adsabs.harvard.edu/abs/2003PASP..115..763C}
}

@ARTICLE{Hatsukade18,
       author = {{Hatsukade}, Bunyo and {Kohno}, Kotaro and {Yamaguchi}, Yuki and {Umehata}, Hideki and {Ao}, Yiping and {Aretxaga}, Itziar and {Caputi}, Karina I. and {Dunlop}, James S. and {Egami}, Eiichi and {Espada}, Daniel and {Fujimoto}, Seiji and {Hayatsu}, Natsuki H. and {Hughes}, David H. and {Ikarashi}, Soh and {Iono}, Daisuke and {Ivison}, Rob J. and {Kawabe}, Ryohei and {Kodama}, Tadayuki and {Lee}, Minju and {Matsuda}, Yuichi and {Nakanishi}, Kouichiro and {Ohta}, Kouji and {Ouchi}, Masami and {Rujopakarn}, Wiphu and {Suzuki}, Tomoko and {Tamura}, Yoichi and {Ueda}, Yoshihiro and {Wang}, Tao and {Wang}, Wei-Hao and {Wilson}, Grant W. and {Yoshimura}, Yuki and {Yun}, Min S.},
        title = "{ALMA twenty-six arcmin$^{2}$ survey of GOODS-S at one millimeter (ASAGAO): Source catalog and number counts}",
      journal = {\pasj},
         year = 2018,
        month = dec,
       volume = {70},
       number = {6},
          eid = {105},
        pages = {105},
          doi = {10.1093/pasj/psy104},
archivePrefix = {arXiv},
       eprint = {1808.04502},
 primaryClass = {astro-ph.GA},
       adsurl = {https://ui.adsabs.harvard.edu/abs/2018PASJ...70..105H}
}

@software{Robitaille12,
       author = {{Robitaille}, Thomas and {Bressert}, Eli},
        title = "{APLpy: Astronomical Plotting Library in Python}",
 howpublished = {Astrophysics Source Code Library, record ascl:1208.017},
         year = 2012,
        month = aug,
          eid = {ascl:1208.017},
       adsurl = {https://ui.adsabs.harvard.edu/abs/2012ascl.soft08017R}
}

@INPROCEEDINGS{Taylor05,
       author = {{Taylor}, M.~B.},
        title = "{TOPCAT \& STIL: Starlink Table/VOTable Processing Software}",
    booktitle = {Astronomical Data Analysis Software and Systems XIV},
         year = 2005,
       editor = {{Shopbell}, P. and {Britton}, M. and {Ebert}, R.},
       series = {Astronomical Society of the Pacific Conference Series},
       volume = {347},
        month = dec,
        pages = {29},
       adsurl = {https://ui.adsabs.harvard.edu/abs/2005ASPC..347...29T}
}

@ARTICLE{Astropy13,
       author = {{Astropy Collaboration} and {Robitaille}, Thomas P. and {Tollerud}, Erik J. and {Greenfield}, Perry and {Droettboom}, Michael and {Bray}, Erik and {Aldcroft}, Tom and {Davis}, Matt and {Ginsburg}, Adam and {Price-Whelan}, Adrian M. and {Kerzendorf}, Wolfgang E. and {Conley}, Alexander and {Crighton}, Neil and {Barbary}, Kyle and {Muna}, Demitri and {Ferguson}, Henry and {Grollier}, Fr{\'e}d{\'e}ric and {Parikh}, Madhura M. and {Nair}, Prasanth H. and {Unther}, Hans M. and {Deil}, Christoph and {Woillez}, Julien and {Conseil}, Simon and {Kramer}, Roban and {Turner}, James E.~H. and {Singer}, Leo and {Fox}, Ryan and {Weaver}, Benjamin A. and {Zabalza}, Victor and {Edwards}, Zachary I. and {Azalee Bostroem}, K. and {Burke}, D.~J. and {Casey}, Andrew R. and {Crawford}, Steven M. and {Dencheva}, Nadia and {Ely}, Justin and {Jenness}, Tim and {Labrie}, Kathleen and {Lim}, Pey Lian and {Pierfederici}, Francesco and {Pontzen}, Andrew and {Ptak}, Andy and {Refsdal}, Brian and {Servillat}, Mathieu and {Streicher}, Ole},
        title = "{Astropy: A community Python package for astronomy}",
      journal = {\aap},
         year = 2013,
        month = oct,
       volume = {558},
          eid = {A33},
        pages = {A33},
          doi = {10.1051/0004-6361/201322068},
archivePrefix = {arXiv},
       eprint = {1307.6212},
 primaryClass = {astro-ph.IM},
       adsurl = {https://ui.adsabs.harvard.edu/abs/2013A&A...558A..33A}
}

@ARTICLE{Virtanen20,
       author = {{Virtanen}, Pauli and {Gommers}, Ralf and {Oliphant}, Travis E. and {Haberland}, Matt and {Reddy}, Tyler and {Cournapeau}, David and {Burovski}, Evgeni and {Peterson}, Pearu and {Weckesser}, Warren and {Bright}, Jonathan and {van der Walt}, St{\'e}fan J. and {Brett}, Matthew and {Wilson}, Joshua and {Millman}, K. Jarrod and {Mayorov}, Nikolay and {Nelson}, Andrew R.~J. and {Jones}, Eric and {Kern}, Robert and {Larson}, Eric and {Carey}, C.~J. and {Polat}, {\.I}lhan and {Feng}, Yu and {Moore}, Eric W. and {VanderPlas}, Jake and {Laxalde}, Denis and {Perktold}, Josef and {Cimrman}, Robert and {Henriksen}, Ian and {Quintero}, E.~A. and {Harris}, Charles R. and {Archibald}, Anne M. and {Ribeiro}, Ant{\^o}nio H. and {Pedregosa}, Fabian and {van Mulbregt}, Paul and {SciPy 1. 0 Contributors}},
        title = "{SciPy 1.0: fundamental algorithms for scientific computing in Python}",
      journal = {Nature Methods},
         year = 2020,
        month = feb,
       volume = {17},
        pages = {261-272},
          doi = {10.1038/s41592-019-0686-2},
archivePrefix = {arXiv},
       eprint = {1907.10121},
 primaryClass = {cs.MS},
       adsurl = {https://ui.adsabs.harvard.edu/abs/2020NatMe..17..261V}
}

@ARTICLE{Hunter07,
  author={Hunter, John D.},
  journal={Computing in Science \& Engineering}, 
  title={Matplotlib: A 2D Graphics Environment}, 
  year={2007},
  volume={9},
  number={3},
  pages={90-95},
  doi={10.1109/MCSE.2007.55}}

@ARTICLE{Walt11,
  author={van der Walt, Stefan and Colbert, S. Chris and Varoquaux, Gael},
  journal={Computing in Science \& Engineering}, 
  title={The NumPy Array: A Structure for Efficient Numerical Computation}, 
  year={2011},
  volume={13},
  number={2},
  pages={22-30},
  doi={10.1109/MCSE.2011.37}}

@ARTICLE{PerezGranger07,
  author={Perez, Fernando and Granger, Brian E.},
  journal={Computing in Science \& Engineering}, 
  title={IPython: A System for Interactive Scientific Computing}, 
  year={2007},
  volume={9},
  number={3},
  pages={21-29},
  doi={10.1109/MCSE.2007.53}}

@ARTICLE{Casertano00,
       author = {{Casertano}, Stefano and {de Mello}, Du{\'\i}lia and {Dickinson}, Mark and {Ferguson}, Henry C. and {Fruchter}, Andrew S. and {Gonzalez-Lopezlira}, Rosa A. and {Heyer}, Inge and {Hook}, Richard N. and {Levay}, Zolt and {Lucas}, Ray A. and {Mack}, Jennifer and {Makidon}, Russell B. and {Mutchler}, Max and {Smith}, T. Ed and {Stiavelli}, Massimo and {Wiggs}, Michael S. and {Williams}, Robert E.},
        title = "{WFPC2 Observations of the Hubble Deep Field South}",
      journal = {\aj},
         year = 2000,
        month = dec,
       volume = {120},
       number = {6},
        pages = {2747-2824},
          doi = {10.1086/316851},
archivePrefix = {arXiv},
       eprint = {astro-ph/0010245},
 primaryClass = {astro-ph},
       adsurl = {https://ui.adsabs.harvard.edu/abs/2000AJ....120.2747C}
}

@ARTICLE{Berta13,
       author = {{Berta}, S. and {Lutz}, D. and {Nordon}, R. and {Genzel}, R. and {Magnelli}, B. and {Popesso}, P. and {Rosario}, D. and {Saintonge}, A. and {Wuyts}, S. and {Tacconi}, L.~J.},
        title = "{Molecular gas mass functions of normal star-forming galaxies since z \raisebox{-0.5ex}\textasciitilde 3}",
      journal = {\aap},
         year = 2013,
        month = jul,
       volume = {555},
          eid = {L8},
        pages = {L8},
          doi = {10.1051/0004-6361/201321776},
archivePrefix = {arXiv},
       eprint = {1304.7771},
 primaryClass = {astro-ph.CO},
       adsurl = {https://ui.adsabs.harvard.edu/abs/2013A&A...555L...8B}
}

@ARTICLE{Fossati19,
       author = {{Fossati}, M. and {Fumagalli}, M. and {Lofthouse}, E.~K. and {D'Odorico}, V. and {Lusso}, E. and {Cantalupo}, S. and {Cooke}, R.~J. and {Cristiani}, S. and {Haardt}, F. and {Morris}, S.~L. and {Peroux}, C. and {Prichard}, L.~J. and {Rafelski}, M. and {Smail}, I. and {Theuns}, T.},
        title = "{The MUSE Ultra Deep Field (MUDF). II. Survey design and the gaseous properties of galaxy groups at 0.5 < z < 1.5}",
      journal = {\mnras},
         year = 2019,
        month = nov,
       volume = {490},
       number = {1},
        pages = {1451-1469},
          doi = {10.1093/mnras/stz2693},
archivePrefix = {arXiv},
       eprint = {1909.04672},
 primaryClass = {astro-ph.GA},
       adsurl = {https://ui.adsabs.harvard.edu/abs/2019MNRAS.490.1451F}
}

@ARTICLE{Bacon17,
       author = {{Bacon}, Roland and {Conseil}, Simon and {Mary}, David and {Brinchmann}, Jarle and {Shepherd}, Martin and {Akhlaghi}, Mohammad and {Weilbacher}, Peter M. and {Piqueras}, Laure and {Wisotzki}, Lutz and {Lagattuta}, David and {Epinat}, Benoit and {Guerou}, Adrien and {Inami}, Hanae and {Cantalupo}, Sebastiano and {Courbot}, Jean Baptiste and {Contini}, Thierry and {Richard}, Johan and {Maseda}, Michael and {Bouwens}, Rychard and {Bouch{\'e}}, Nicolas and {Kollatschny}, Wolfram and {Schaye}, Joop and {Marino}, Raffaella Anna and {Pello}, Roser and {Herenz}, Christian and {Guiderdoni}, Bruno and {Carollo}, Marcella},
        title = "{The MUSE Hubble Ultra Deep Field Survey. I. Survey description, data reduction, and source detection}",
      journal = {\aap},
         year = 2017,
        month = dec,
       volume = {608},
          eid = {A1},
        pages = {A1},
          doi = {10.1051/0004-6361/201730833},
archivePrefix = {arXiv},
       eprint = {1710.03002},
 primaryClass = {astro-ph.GA},
       adsurl = {https://ui.adsabs.harvard.edu/abs/2017A&A...608A...1B}
}

@ARTICLE{Gaia23,
       author = {{Gaia Collaboration} and {Vallenari}, A. and {Brown}, A.~G.~A. and {Prusti}, T. and {de Bruijne}, J.~H.~J. and {Arenou}, F. and {Babusiaux}, C. and {Biermann}, M. and {Creevey}, O.~L. and {Ducourant}, C. and {Evans}, D.~W. and {Eyer}, L. and {Guerra}, R. and {Hutton}, A. and {Jordi}, C. and {Klioner}, S.~A. and {Lammers}, U.~L. and {Lindegren}, L. and {Luri}, X. and {Mignard}, F. and {Panem}, C. and {Pourbaix}, D. and {Randich}, S. and {Sartoretti}, P. and {Soubiran}, C. and {Tanga}, P. and {Walton}, N.~A. and {Bailer-Jones}, C.~A.~L. and {Bastian}, U. and {Drimmel}, R. and {Jansen}, F. and {Katz}, D. and {Lattanzi}, M.~G. and {van Leeuwen}, F. and {Bakker}, J. and {Cacciari}, C. and {Casta{\~n}eda}, J. and {De Angeli}, F. and {Fabricius}, C. and {Fouesneau}, M. and {Fr{\'e}mat}, Y. and {Galluccio}, L. and {Guerrier}, A. and {Heiter}, U. and {Masana}, E. and {Messineo}, R. and {Mowlavi}, N. and {Nicolas}, C. and {Nienartowicz}, K. and {Pailler}, F. and {Panuzzo}, P. and {Riclet}, F. and {Roux}, W. and {Seabroke}, G.~M. and {Sordo}, R. and {Th{\'e}venin}, F. and {Gracia-Abril}, G. and {Portell}, J. and {Teyssier}, D. and {Altmann}, M. and {Andrae}, R. and {Audard}, M. and {Bellas-Velidis}, I. and {Benson}, K. and {Berthier}, J. and {Blomme}, R. and {Burgess}, P.~W. and {Busonero}, D. and {Busso}, G. and {C{\'a}novas}, H. and {Carry}, B. and {Cellino}, A. and {Cheek}, N. and {Clementini}, G. and {Damerdji}, Y. and {Davidson}, M. and {de Teodoro}, P. and {Nu{\~n}ez Campos}, M. and {Delchambre}, L. and {Dell'Oro}, A. and {Esquej}, P. and {Fern{\'a}ndez-Hern{\'a}ndez}, J. and {Fraile}, E. and {Garabato}, D. and {Garc{\'\i}a-Lario}, P. and {Gosset}, E. and {Haigron}, R. and {Halbwachs}, J. -L. and {Hambly}, N.~C. and {Harrison}, D.~L. and {Hern{\'a}ndez}, J. and {Hestroffer}, D. and {Hodgkin}, S.~T. and {Holl}, B. and {Jan{\ss}en}, K. and {Jevardat de Fombelle}, G. and {Jordan}, S. and {Krone-Martins}, A. and {Lanzafame}, A.~C. and {L{\"o}ffler}, W. and {Marchal}, O. and {Marrese}, P.~M. and {Moitinho}, A. and {Muinonen}, K. and {Osborne}, P. and {Pancino}, E. and {Pauwels}, T. and {Recio-Blanco}, A. and {Reyl{\'e}}, C. and {Riello}, M. and {Rimoldini}, L. and {Roegiers}, T. and {Rybizki}, J. and {Sarro}, L.~M. and {Siopis}, C. and {Smith}, M. and {Sozzetti}, A. and {Utrilla}, E. and {van Leeuwen}, M. and {Abbas}, U. and {{\'A}brah{\'a}m}, P. and {Abreu Aramburu}, A. and {Aerts}, C. and {Aguado}, J.~J. and {Ajaj}, M. and {Aldea-Montero}, F. and {Altavilla}, G. and {{\'A}lvarez}, M.~A. and {Alves}, J. and {Anders}, F. and {Anderson}, R.~I. and {Anglada Varela}, E. and {Antoja}, T. and {Baines}, D. and {Baker}, S.~G. and {Balaguer-N{\'u}{\~n}ez}, L. and {Balbinot}, E. and {Balog}, Z. and {Barache}, C. and {Barbato}, D. and {Barros}, M. and {Barstow}, M.~A. and {Bartolom{\'e}}, S. and {Bassilana}, J. -L. and {Bauchet}, N. and {Becciani}, U. and {Bellazzini}, M. and {Berihuete}, A. and {Bernet}, M. and {Bertone}, S. and {Bianchi}, L. and {Binnenfeld}, A. and {Blanco-Cuaresma}, S. and {Blazere}, A. and {Boch}, T. and {Bombrun}, A. and {Bossini}, D. and {Bouquillon}, S. and {Bragaglia}, A. and {Bramante}, L. and {Breedt}, E. and {Bressan}, A. and {Brouillet}, N. and {Brugaletta}, E. and {Bucciarelli}, B. and {Burlacu}, A. and {Butkevich}, A.~G. and {Buzzi}, R. and {Caffau}, E. and {Cancelliere}, R. and {Cantat-Gaudin}, T. and {Carballo}, R. and {Carlucci}, T. and {Carnerero}, M.~I. and {Carrasco}, J.~M. and {Casamiquela}, L. and {Castellani}, M. and {Castro-Ginard}, A. and {Chaoul}, L. and {Charlot}, P. and {Chemin}, L. and {Chiaramida}, V. and {Chiavassa}, A. and {Chornay}, N. and {Comoretto}, G. and {Contursi}, G. and {Cooper}, W.~J. and {Cornez}, T. and {Cowell}, S. and {Crifo}, F. and {Cropper}, M. and {Crosta}, M. and {Crowley}, C. and {Dafonte}, C. and {Dapergolas}, A. and {David}, M. and {David}, P. and {de Laverny}, P. and {De Luise}, F. and {De March}, R.},
        title = "{Gaia Data Release 3. Summary of the content and survey properties}",
      journal = {\aap},
         year = 2023,
        month = jun,
       volume = {674},
          eid = {A1},
        pages = {A1},
          doi = {10.1051/0004-6361/202243940},
archivePrefix = {arXiv},
       eprint = {2208.00211},
 primaryClass = {astro-ph.GA},
       adsurl = {https://ui.adsabs.harvard.edu/abs/2023A&A...674A...1G}
}

@ARTICLE{Contini16,
       author = {{Contini}, T. and {Epinat}, B. and {Bouch{\'e}}, N. and {Brinchmann}, J. and {Boogaard}, L.~A. and {Ventou}, E. and {Bacon}, R. and {Richard}, J. and {Weilbacher}, P.~M. and {Wisotzki}, L. and {Krajnovi{\'c}}, D. and {Vielfaure}, J. -B. and {Emsellem}, E. and {Finley}, H. and {Inami}, H. and {Schaye}, J. and {Swinbank}, M. and {Gu{\'e}rou}, A. and {Martinsson}, T. and {Michel-Dansac}, L. and {Schroetter}, I. and {Shirazi}, M. and {Soucail}, G.},
        title = "{Deep MUSE observations in the HDFS. Morpho-kinematics of distant star-forming galaxies down to {}10$^{8}$M$_{{\ensuremath{\odot}}}$}",
      journal = {\aap},
         year = 2016,
        month = jun,
       volume = {591},
          eid = {A49},
        pages = {A49},
          doi = {10.1051/0004-6361/201527866},
archivePrefix = {arXiv},
       eprint = {1512.00246},
 primaryClass = {astro-ph.GA},
       adsurl = {https://ui.adsabs.harvard.edu/abs/2016A&A...591A..49C}
}

@ARTICLE{Kneissl19,
       author = {{Kneissl}, R{\"u}diger and {Polletta}, Maria del Carmen and {Martinache}, Clement and {Hill}, Ryley and {Clarenc}, Benjamin and {Dole}, Herve A. and {Nesvadba}, Nicole P.~H. and {Scott}, Douglas and {B{\'e}thermin}, Matthieu and {Frye}, Brenda and {Giard}, Martin and {Lagache}, Guilaine and {Montier}, Ludovic},
        title = "{Using ALMA to resolve the nature of the early star-forming large-scale structure PLCK G073.4-57.5}",
      journal = {\aap},
         year = 2019,
        month = may,
       volume = {625},
          eid = {A96},
        pages = {A96},
          doi = {10.1051/0004-6361/201833252},
archivePrefix = {arXiv},
       eprint = {1804.06581},
 primaryClass = {astro-ph.GA},
       adsurl = {https://ui.adsabs.harvard.edu/abs/2019A&A...625A..96K}
}

@ARTICLE{Hill25,
       author = {{Hill}, Ryley and {Polletta}, Maria del Carmen and {B{\'e}thermin}, Matthieu and {Dole}, Herv{\'e} and {Kneissl}, R{\"u}diger and {Scott}, Douglas},
        title = "{An ALMA spectroscopic survey of the Planck high-redshift object PLCK G073.4‑57.5 confirms two protoclusters}",
      journal = {\aap},
         year = 2025,
        month = jun,
       volume = {698},
          eid = {A204},
        pages = {A204},
          doi = {10.1051/0004-6361/202453238},
archivePrefix = {arXiv},
       eprint = {2412.00294},
 primaryClass = {astro-ph.CO},
       adsurl = {https://ui.adsabs.harvard.edu/abs/2025A&A...698A.204H}
}

@ARTICLE{Drake17,
       author = {{Drake}, A.~B. and {Garel}, T. and {Wisotzki}, L. and {Leclercq}, F. and {Hashimoto}, T. and {Richard}, J. and {Bacon}, R. and {Blaizot}, J. and {Caruana}, J. and {Conseil}, S. and {Contini}, T. and {Guiderdoni}, B. and {Herenz}, E.~C. and {Inami}, H. and {Lewis}, J. and {Mahler}, G. and {Marino}, R.~A. and {Pello}, R. and {Schaye}, J. and {Verhamme}, A. and {Ventou}, E. and {Weilbacher}, P.~M.},
        title = "{The MUSE Hubble Ultra Deep Field Survey. VI. The faint-end of the Ly{\ensuremath{\alpha}} luminosity function at 2.91 < z < 6.64 and implications for reionisation}",
      journal = {\aap},
         year = 2017,
        month = nov,
       volume = {608},
          eid = {A6},
        pages = {A6},
          doi = {10.1051/0004-6361/201731431},
archivePrefix = {arXiv},
       eprint = {1711.03095},
 primaryClass = {astro-ph.CO},
       adsurl = {https://ui.adsabs.harvard.edu/abs/2017A&A...608A...6D}
}

@ARTICLE{Huynh05,
       author = {{Huynh}, Minh T. and {Jackson}, Carole A. and {Norris}, Ray P. and {Prandoni}, Isabella},
        title = "{Radio Observations of the Hubble Deep Field-South Region. II. The 1.4 GHz Catalog and Source Counts}",
      journal = {\aj},
         year = 2005,
        month = oct,
       volume = {130},
       number = {4},
        pages = {1373-1388},
          doi = {10.1086/432873},
archivePrefix = {arXiv},
       eprint = {astro-ph/0506047},
 primaryClass = {astro-ph},
       adsurl = {https://ui.adsabs.harvard.edu/abs/2005AJ....130.1373H}
}

@ARTICLE{Huynh07,
       author = {{Huynh}, Minh T. and {Jackson}, Carole A. and {Norris}, Ray P.},
        title = "{Radio Observations of the Hubble Deep Field-South Region. III. The 2.5, 5.2, and 8.7 GHz Catalogs and Radio Source Properties}",
      journal = {\aj},
         year = 2007,
        month = apr,
       volume = {133},
       number = {4},
        pages = {1331-1344},
          doi = {10.1086/511420},
archivePrefix = {arXiv},
       eprint = {astro-ph/0612518},
 primaryClass = {astro-ph},
       adsurl = {https://ui.adsabs.harvard.edu/abs/2007AJ....133.1331H}
}

@ARTICLE{Wilson11,
       author = {{Wilson}, Warwick E. and {Ferris}, R.~H. and {Axtens}, P. and {Brown}, A. and {Davis}, E. and {Hampson}, G. and {Leach}, M. and {Roberts}, P. and {Saunders}, S. and {Koribalski}, B.~S. and {Caswell}, J.~L. and {Lenc}, E. and {Stevens}, J. and {Voronkov}, M.~A. and {Wieringa}, M.~H. and {Brooks}, K. and {Edwards}, P.~G. and {Ekers}, R.~D. and {Emonts}, B. and {Hindson}, L. and {Johnston}, S. and {Maddison}, S.~T. and {Mahony}, E.~K. and {Malu}, S.~S. and {Massardi}, M. and {Mao}, M.~Y. and {McConnell}, D. and {Norris}, R.~P. and {Schnitzeler}, D. and {Subrahmanyan}, R. and {Urquhart}, J.~S. and {Thompson}, M.~A. and {Wark}, R.~M.},
        title = "{The Australia Telescope Compact Array Broad-band Backend: description and first results}",
      journal = {\mnras},
         year = 2011,
        month = sep,
       volume = {416},
       number = {2},
        pages = {832-856},
          doi = {10.1111/j.1365-2966.2011.19054.x},
archivePrefix = {arXiv},
       eprint = {1105.3532},
 primaryClass = {astro-ph.IM},
       adsurl = {https://ui.adsabs.harvard.edu/abs/2011MNRAS.416..832W}
}

@ARTICLE{Schmidt68,
       author = {{Schmidt}, Maarten},
        title = "{Space Distribution and Luminosity Functions of Quasi-Stellar Radio Sources}",
      journal = {\apj},
         year = 1968,
        month = feb,
       volume = {151},
        pages = {393},
          doi = {10.1086/149446},
       adsurl = {https://ui.adsabs.harvard.edu/abs/1968ApJ...151..393S}
}

@ARTICLE{Scoville17,
       author = {{Scoville}, N. and {Lee}, N. and {Vanden Bout}, P. and {Diaz-Santos}, T. and {Sanders}, D. and {Darvish}, B. and {Bongiorno}, A. and {Casey}, C.~M. and {Murchikova}, L. and {Koda}, J. and {Capak}, P. and {Vlahakis}, Catherine and {Ilbert}, O. and {Sheth}, K. and {Morokuma-Matsui}, K. and {Ivison}, R.~J. and {Aussel}, H. and {Laigle}, C. and {McCracken}, H.~J. and {Armus}, L. and {Pope}, A. and {Toft}, S. and {Masters}, D.},
        title = "{Evolution of Interstellar Medium, Star Formation, and Accretion at High Redshift}",
      journal = {\apj},
         year = 2017,
        month = mar,
       volume = {837},
       number = {2},
          eid = {150},
        pages = {150},
          doi = {10.3847/1538-4357/aa61a0},
archivePrefix = {arXiv},
       eprint = {1702.04729},
 primaryClass = {astro-ph.GA},
       adsurl = {https://ui.adsabs.harvard.edu/abs/2017ApJ...837..150S}
}

@ARTICLE{Messias24,
       author = {{Messias}, Hugo and {Guerrero}, Andrea and {Nagar}, Neil and {Regueiro}, Jack and {Impellizzeri}, Violette and {Orellana}, Gustavo and {Vioque}, Miguel},
        title = "{H I content at cosmic noon - a millimetre-wavelength perspective}",
      journal = {\mnras},
         year = 2024,
        month = oct,
       volume = {533},
       number = {4},
        pages = {3937-3956},
          doi = {10.1093/mnras/stae1807},
archivePrefix = {arXiv},
       eprint = {2312.02782},
 primaryClass = {astro-ph.GA},
       adsurl = {https://ui.adsabs.harvard.edu/abs/2024MNRAS.533.3937M}
}

@ARTICLE{TrentiStiavelli08,
       author = {{Trenti}, M. and {Stiavelli}, M.},
        title = "{Cosmic Variance and Its Effect on the Luminosity Function Determination in Deep High-z Surveys}",
      journal = {\apj},
         year = 2008,
        month = apr,
       volume = {676},
       number = {2},
        pages = {767-780},
          doi = {10.1086/528674},
archivePrefix = {arXiv},
       eprint = {0712.0398},
 primaryClass = {astro-ph},
       adsurl = {https://ui.adsabs.harvard.edu/abs/2008ApJ...676..767T}
}

@ARTICLE{Decarli20,
       author = {{Decarli}, Roberto and {Aravena}, Manuel and {Boogaard}, Leindert and {Carilli}, Chris and {Gonz{\'a}lez-L{\'o}pez}, Jorge and {Walter}, Fabian and {Cortes}, Paulo C. and {Cox}, Pierre and {da Cunha}, Elisabete and {Daddi}, Emanuele and {D{\'\i}az-Santos}, Tanio and {Hodge}, Jacqueline A. and {Inami}, Hanae and {Neeleman}, Marcel and {Novak}, Mladen and {Oesch}, Pascal and {Popping}, Gerg{\"o} and {Riechers}, Dominik and {Smail}, Ian and {Uzgil}, Bade and {van der Werf}, Paul and {Wagg}, Jeff and {Weiss}, Axel},
        title = "{The ALMA Spectroscopic Survey in the Hubble Ultra Deep Field: Multiband Constraints on Line-luminosity Functions and the Cosmic Density of Molecular Gas}",
      journal = {\apj},
         year = 2020,
        month = oct,
       volume = {902},
       number = {2},
          eid = {110},
        pages = {110},
          doi = {10.3847/1538-4357/abaa3b},
archivePrefix = {arXiv},
       eprint = {2009.10744},
 primaryClass = {astro-ph.GA},
       adsurl = {https://ui.adsabs.harvard.edu/abs/2020ApJ...902..110D}
}

@ARTICLE{Kaasinen17,
       author = {{Kaasinen}, Melanie and {Bian}, Fuyan and {Groves}, Brent and {Kewley}, Lisa J. and {Gupta}, Anshu},
        title = "{The COSMOS-[O II] survey: evolution of electron density with star formation rate}",
      journal = {\mnras},
         year = 2017,
        month = mar,
       volume = {465},
       number = {3},
        pages = {3220-3234},
          doi = {10.1093/mnras/stw2827},
archivePrefix = {arXiv},
       eprint = {1611.01166},
 primaryClass = {astro-ph.GA},
       adsurl = {https://ui.adsabs.harvard.edu/abs/2017MNRAS.465.3220K}
}

@INPROCEEDINGS{Bacon10,
       author = {{Bacon}, R. and {Accardo}, M. and {Adjali}, L. and {Anwand}, H. and {Bauer}, S. and {Biswas}, I. and {Blaizot}, J. and {Boudon}, D. and {Brau-Nogue}, S. and {Brinchmann}, J. and {Caillier}, P. and {Capoani}, L. and {Carollo}, C.~M. and {Contini}, T. and {Couderc}, P. and {Daguis{\'e}}, E. and {Deiries}, S. and {Delabre}, B. and {Dreizler}, S. and {Dubois}, J. and {Dupieux}, M. and {Dupuy}, C. and {Emsellem}, E. and {Fechner}, T. and {Fleischmann}, A. and {Fran{\c{c}}ois}, M. and {Gallou}, G. and {Gharsa}, T. and {Glindemann}, A. and {Gojak}, D. and {Guiderdoni}, B. and {Hansali}, G. and {Hahn}, T. and {Jarno}, A. and {Kelz}, A. and {Koehler}, C. and {Kosmalski}, J. and {Laurent}, F. and {Le Floch}, M. and {Lilly}, S.~J. and {Lizon}, J. -L. and {Loupias}, M. and {Manescau}, A. and {Monstein}, C. and {Nicklas}, H. and {Olaya}, J. -C. and {Pares}, L. and {Pasquini}, L. and {P{\'e}contal-Rousset}, A. and {Pell{\'o}}, R. and {Petit}, C. and {Popow}, E. and {Reiss}, R. and {Remillieux}, A. and {Renault}, E. and {Roth}, M. and {Rupprecht}, G. and {Serre}, D. and {Schaye}, J. and {Soucail}, G. and {Steinmetz}, M. and {Streicher}, O. and {Stuik}, R. and {Valentin}, H. and {Vernet}, J. and {Weilbacher}, P. and {Wisotzki}, L. and {Yerle}, N.},
        title = "{The MUSE second-generation VLT instrument}",
    booktitle = {Ground-based and Airborne Instrumentation for Astronomy III},
         year = 2010,
       editor = {{McLean}, Ian S. and {Ramsay}, Suzanne K. and {Takami}, Hideki},
       series = {Society of Photo-Optical Instrumentation Engineers (SPIE) Conference Series},
       volume = {7735},
        month = jul,
          eid = {773508},
        pages = {773508},
          doi = {10.1117/12.856027},
archivePrefix = {arXiv},
       eprint = {2211.16795},
 primaryClass = {astro-ph.IM},
       adsurl = {https://ui.adsabs.harvard.edu/abs/2010SPIE.7735E..08B}
}

@BOOK{vltwb,
       author = {{European Southern Observatory}},
        title = "{The VLT White Book}",
         year = 1998,
       adsurl = {https://ui.adsabs.harvard.edu/abs/1998vltw.book.....E}
}

@ARTICLE{Decarli16,
       author = {{Decarli}, Roberto and {Walter}, Fabian and {Aravena}, Manuel and {Carilli}, Chris and {Bouwens}, Rychard and {da Cunha}, Elisabete and {Daddi}, Emanuele and {Ivison}, R.~J. and {Popping}, Gerg{\"o} and {Riechers}, Dominik and {Smail}, Ian R. and {Swinbank}, Mark and {Weiss}, Axel and {Anguita}, Timo and {Assef}, Roberto J. and {Bauer}, Franz E. and {Bell}, Eric F. and {Bertoldi}, Frank and {Chapman}, Scott and {Colina}, Luis and {Cortes}, Paulo C. and {Cox}, Pierre and {Dickinson}, Mark and {Elbaz}, David and {G{\'o}nzalez-L{\'o}pez}, Jorge and {Ibar}, Edo and {Infante}, Leopoldo and {Hodge}, Jacqueline and {Karim}, Alex and {Le Fevre}, Olivier and {Magnelli}, Benjamin and {Neri}, Roberto and {Oesch}, Pascal and {Ota}, Kazuaki and {Rix}, Hans-Walter and {Sargent}, Mark and {Sheth}, Kartik and {van der Wel}, Arjen and {van der Werf}, Paul and {Wagg}, Jeff},
        title = "{ALMA Spectroscopic Survey in the Hubble Ultra Deep Field: CO Luminosity Functions and the Evolution of the Cosmic Density of Molecular Gas}",
      journal = {\apj},
         year = 2016,
        month = dec,
       volume = {833},
       number = {1},
          eid = {69},
        pages = {69},
          doi = {10.3847/1538-4357/833/1/69},
archivePrefix = {arXiv},
       eprint = {1607.06770},
 primaryClass = {astro-ph.GA},
       adsurl = {https://ui.adsabs.harvard.edu/abs/2016ApJ...833...69D}
}

@ARTICLE{Decarli19,
       author = {{Decarli}, Roberto and {Walter}, Fabian and {G{\'o}nzalez-L{\'o}pez}, Jorge and {Aravena}, Manuel and {Boogaard}, Leindert and {Carilli}, Chris and {Cox}, Pierre and {Daddi}, Emanuele and {Popping}, Gerg{\"o} and {Riechers}, Dominik and {Uzgil}, Bade and {Weiss}, Axel and {Assef}, Roberto J. and {Bacon}, Roland and {Bauer}, Franz Erik and {Bertoldi}, Frank and {Bouwens}, Rychard and {Contini}, Thierry and {Cortes}, Paulo C. and {da Cunha}, Elisabete and {D{\'\i}az-Santos}, Tanio and {Elbaz}, David and {Inami}, Hanae and {Hodge}, Jacqueline and {Ivison}, Rob and {Le F{\`e}vre}, Olivier and {Magnelli}, Benjamin and {Novak}, Mladen and {Oesch}, Pascal and {Rix}, Hans-Walter and {Sargent}, Mark T. and {Smail}, Ian and {Swinbank}, A. Mark and {Somerville}, Rachel S. and {van der Werf}, Paul and {Wagg}, Jeff and {Wisotzki}, Lutz},
        title = "{The ALMA Spectroscopic Survey in the HUDF: CO Luminosity Functions and the Molecular Gas Content of Galaxies through Cosmic History}",
      journal = {\apj},
         year = 2019,
        month = sep,
       volume = {882},
       number = {2},
          eid = {138},
        pages = {138},
          doi = {10.3847/1538-4357/ab30fe},
archivePrefix = {arXiv},
       eprint = {1903.09164},
 primaryClass = {astro-ph.GA},
       adsurl = {https://ui.adsabs.harvard.edu/abs/2019ApJ...882..138D}
}

@ARTICLE{Oteo18,
       author = {{Oteo}, I. and {Ivison}, R.~J. and {Dunne}, L. and {Manilla-Robles}, A. and {Maddox}, S. and {Lewis}, A.~J.~R. and {de Zotti}, G. and {Bremer}, M. and {Clements}, D.~L. and {Cooray}, A. and {Dannerbauer}, H. and {Eales}, S. and {Greenslade}, J. and {Omont}, A. and {Perez{\textendash}Fourn{\'o}n}, I. and {Riechers}, D. and {Scott}, D. and {van der Werf}, P. and {Weiss}, A. and {Zhang}, Z. -Y.},
        title = "{An Extreme Protocluster of Luminous Dusty Starbursts in the Early Universe}",
      journal = {\apj},
         year = 2018,
        month = mar,
       volume = {856},
       number = {1},
          eid = {72},
        pages = {72},
          doi = {10.3847/1538-4357/aaa1f1},
archivePrefix = {arXiv},
       eprint = {1709.02809},
 primaryClass = {astro-ph.GA},
       adsurl = {https://ui.adsabs.harvard.edu/abs/2018ApJ...856...72O}
}

@ARTICLE{Miller18,
       author = {{Miller}, T.~B. and {Chapman}, S.~C. and {Aravena}, M. and {Ashby}, M.~L.~N. and {Hayward}, C.~C. and {Vieira}, J.~D. and {Wei{\ss}}, A. and {Babul}, A. and {B{\'e}thermin}, M. and {Bradford}, C.~M. and {Brodwin}, M. and {Carlstrom}, J.~E. and {Chen}, Chian-Chou and {Cunningham}, D.~J.~M. and {De Breuck}, C. and {Gonzalez}, A.~H. and {Greve}, T.~R. and {Harnett}, J. and {Hezaveh}, Y. and {Lacaille}, K. and {Litke}, K.~C. and {Ma}, J. and {Malkan}, M. and {Marrone}, D.~P. and {Morningstar}, W. and {Murphy}, E.~J. and {Narayanan}, D. and {Pass}, E. and {Perry}, R. and {Phadke}, K.~A. and {Rennehan}, D. and {Rotermund}, K.~M. and {Simpson}, J. and {Spilker}, J.~S. and {Sreevani}, J. and {Stark}, A.~A. and {Strandet}, M.~L. and {Strom}, A.~L.},
        title = "{A massive core for a cluster of galaxies at a redshift of 4.3}",
      journal = {\nat},
         year = 2018,
        month = apr,
       volume = {556},
       number = {7702},
        pages = {469-472},
          doi = {10.1038/s41586-018-0025-2},
archivePrefix = {arXiv},
       eprint = {1804.09231},
 primaryClass = {astro-ph.GA},
       adsurl = {https://ui.adsabs.harvard.edu/abs/2018Natur.556..469M}
}

@ARTICLE{Vulcani17,
       author = {{Vulcani}, Benedetta and {Treu}, Tommaso and {Nipoti}, Carlo and {Schmidt}, Kasper B. and {Dressler}, Alan and {Morshita}, Takahiro and {Poggianti}, Bianca M. and {Malkan}, Matthew and {Hoag}, Austin and {Brada{\v{c}}}, Marusa and {Abramson}, Louis and {Trenti}, Michele and {Pentericci}, Laura and {von der Linden}, Anja and {Morris}, Glenn and {Wang}, Xin},
        title = "{The Grism Lens-Amplified Survey from Space (GLASS). VIII. The Influence of the Cluster Properties on H{\ensuremath{\alpha}} Emitter Galaxies at 0.3 < z < 0.7}",
      journal = {\apj},
         year = 2017,
        month = mar,
       volume = {837},
       number = {2},
          eid = {126},
        pages = {126},
          doi = {10.3847/1538-4357/aa618b},
archivePrefix = {arXiv},
       eprint = {1610.04615},
 primaryClass = {astro-ph.GA},
       adsurl = {https://ui.adsabs.harvard.edu/abs/2017ApJ...837..126V}
}

@ARTICLE{Vulcani15,
       author = {{Vulcani}, Benedetta and {Poggianti}, Bianca M. and {Fritz}, Jacopo and {Fasano}, Giovanni and {Moretti}, Alessia and {Calvi}, Rosa and {Paccagnella}, Angela},
        title = "{From Blue Star-forming to Red Passive: Galaxies in Transition in Different Environments}",
      journal = {\apj},
         year = 2015,
        month = jan,
       volume = {798},
       number = {1},
          eid = {52},
        pages = {52},
          doi = {10.1088/0004-637X/798/1/52},
archivePrefix = {arXiv},
       eprint = {1410.6481},
 primaryClass = {astro-ph.GA},
       adsurl = {https://ui.adsabs.harvard.edu/abs/2015ApJ...798...52V}
}

@ARTICLE{Bianconi18,
       author = {{Bianconi}, M. and {Smith}, G.~P. and {Haines}, C.~P. and {McGee}, S.~L. and {Finoguenov}, A. and {Egami}, E.},
        title = "{LoCuSS: pre-processing in galaxy groups falling into massive galaxy clusters at z = 0.2}",
      journal = {\mnras},
         year = 2018,
        month = jan,
       volume = {473},
       number = {1},
        pages = {L79-L83},
          doi = {10.1093/mnrasl/slx167},
archivePrefix = {arXiv},
       eprint = {1710.04230},
 primaryClass = {astro-ph.GA},
       adsurl = {https://ui.adsabs.harvard.edu/abs/2018MNRAS.473L..79B}
}

@ARTICLE{Fujita04,
       author = {{Fujita}, Yutaka},
        title = "{Pre-Processing of Galaxies before Entering a Cluster}",
      journal = {\pasj},
         year = 2004,
        month = feb,
       volume = {56},
        pages = {29-43},
          doi = {10.1093/pasj/56.1.29},
archivePrefix = {arXiv},
       eprint = {astro-ph/0311193},
 primaryClass = {astro-ph},
       adsurl = {https://ui.adsabs.harvard.edu/abs/2004PASJ...56...29F}
}

@ARTICLE{BrownWildCunningham04,
       author = {{Brown}, Robert L. and {Wild}, Wolfgang and {Cunningham}, Charles},
        title = "{ALMA - the Atacama large millimeter array}",
      journal = {Advances in Space Research},
         year = 2004,
        month = jan,
       volume = {34},
       number = {3},
        pages = {555-559},
          doi = {10.1016/j.asr.2003.03.028},
       adsurl = {https://ui.adsabs.harvard.edu/abs/2004AdSpR..34..555B}
}

@ARTICLE{Lenkic20,
       author = {{Lenki{\'c}}, Laura and {Bolatto}, Alberto D. and {F{\"o}rster Schreiber}, Natascha M. and {Tacconi}, Linda J. and {Neri}, Roberto and {Combes}, Francoise and {Walter}, Fabian and {Garc{\'\i}a-Burillo}, Santiago and {Genzel}, Reinhard and {Lutz}, Dieter and {Cooper}, Michael C.},
        title = "{Plateau de Bure High-z Blue Sequence Survey 2 (PHIBSS2): Search for Secondary Sources, CO Luminosity Functions in the Field, and the Evolution of Molecular Gas Density through Cosmic Time}",
      journal = {\aj},
         year = 2020,
        month = may,
       volume = {159},
       number = {5},
          eid = {190},
        pages = {190},
          doi = {10.3847/1538-3881/ab7458},
archivePrefix = {arXiv},
       eprint = {1908.01791},
 primaryClass = {astro-ph.GA},
       adsurl = {https://ui.adsabs.harvard.edu/abs/2020AJ....159..190L}
}

@ARTICLE{Riechers19,
       author = {{Riechers}, Dominik A. and {Pavesi}, Riccardo and {Sharon}, Chelsea E. and {Hodge}, Jacqueline A. and {Decarli}, Roberto and {Walter}, Fabian and {Carilli}, Christopher L. and {Aravena}, Manuel and {da Cunha}, Elisabete and {Daddi}, Emanuele and {Dickinson}, Mark and {Smail}, Ian and {Capak}, Peter L. and {Ivison}, Rob J. and {Sargent}, Mark and {Scoville}, Nicholas Z. and {Wagg}, Jeff},
        title = "{COLDz: Shape of the CO Luminosity Function at High Redshift and the Cold Gas History of the Universe}",
      journal = {\apj},
         year = 2019,
        month = feb,
       volume = {872},
       number = {1},
          eid = {7},
        pages = {7},
          doi = {10.3847/1538-4357/aafc27},
archivePrefix = {arXiv},
       eprint = {1808.04371},
 primaryClass = {astro-ph.GA},
       adsurl = {https://ui.adsabs.harvard.edu/abs/2019ApJ...872....7R}
}

@ARTICLE{Schmidt59,
       author = {{Schmidt}, Maarten},
        title = "{The Rate of Star Formation.}",
      journal = {\apj},
         year = 1959,
        month = mar,
       volume = {129},
        pages = {243},
          doi = {10.1086/146614},
       adsurl = {https://ui.adsabs.harvard.edu/abs/1959ApJ...129..243S}
}

@ARTICLE{Kennicutt98,
       author = {{Kennicutt}, Robert C., Jr.},
        title = "{The Global Schmidt Law in Star-forming Galaxies}",
      journal = {\apj},
         year = 1998,
        month = may,
       volume = {498},
       number = {2},
        pages = {541-552},
          doi = {10.1086/305588},
archivePrefix = {arXiv},
       eprint = {astro-ph/9712213},
 primaryClass = {astro-ph},
       adsurl = {https://ui.adsabs.harvard.edu/abs/1998ApJ...498..541K}
}

@ARTICLE{Walter20,
       author = {{Walter}, Fabian and {Carilli}, Chris and {Neeleman}, Marcel and {Decarli}, Roberto and {Popping}, Gerg{\"o} and {Somerville}, Rachel S. and {Aravena}, Manuel and {Bertoldi}, Frank and {Boogaard}, Leindert and {Cox}, Pierre and {da Cunha}, Elisabete and {Magnelli}, Benjamin and {Obreschkow}, Danail and {Riechers}, Dominik and {Rix}, Hans-Walter and {Smail}, Ian and {Weiss}, Axel and {Assef}, Roberto J. and {Bauer}, Franz and {Bouwens}, Rychard and {Contini}, Thierry and {Cortes}, Paulo C. and {Daddi}, Emanuele and {Diaz-Santos}, Tanio and {Gonz{\'a}lez-L{\'o}pez}, Jorge and {Hennawi}, Joseph and {Hodge}, Jacqueline A. and {Inami}, Hanae and {Ivison}, Rob and {Oesch}, Pascal and {Sargent}, Mark and {van der Werf}, Paul and {Wagg}, Jeff and {Yung}, L.~Y. Aaron},
        title = "{The Evolution of the Baryons Associated with Galaxies Averaged over Cosmic Time and Space}",
      journal = {\apj},
         year = 2020,
        month = oct,
       volume = {902},
       number = {2},
          eid = {111},
        pages = {111},
          doi = {10.3847/1538-4357/abb82e},
archivePrefix = {arXiv},
       eprint = {2009.11126},
 primaryClass = {astro-ph.GA},
       adsurl = {https://ui.adsabs.harvard.edu/abs/2020ApJ...902..111W}
}

@ARTICLE{Hill24,
       author = {{Hill}, Ryley and {Scott}, Douglas and {McLeod}, Derek J. and {McLure}, Ross J. and {Chapman}, Scott C. and {Dunlop}, James S.},
        title = "{An optimal ALMA image of the Hubble Ultra Deep Field in the era of JWST: obscured star formation and the cosmic far-infrared background}",
      journal = {\mnras},
         year = 2024,
        month = mar,
       volume = {528},
       number = {3},
        pages = {5019-5045},
          doi = {10.1093/mnras/stae346},
archivePrefix = {arXiv},
       eprint = {2309.10988},
 primaryClass = {astro-ph.CO},
       adsurl = {https://ui.adsabs.harvard.edu/abs/2024MNRAS.528.5019H}
}

@ARTICLE{Walter16,
       author = {{Walter}, Fabian and {Decarli}, Roberto and {Aravena}, Manuel and {Carilli}, Chris and {Bouwens}, Rychard and {da Cunha}, Elisabete and {Daddi}, Emanuele and {Ivison}, R.~J. and {Riechers}, Dominik and {Smail}, Ian and {Swinbank}, Mark and {Weiss}, Axel and {Anguita}, Timo and {Assef}, Roberto and {Bacon}, Roland and {Bauer}, Franz and {Bell}, Eric F. and {Bertoldi}, Frank and {Chapman}, Scott and {Colina}, Luis and {Cortes}, Paulo C. and {Cox}, Pierre and {Dickinson}, Mark and {Elbaz}, David and {G{\'o}nzalez-L{\'o}pez}, Jorge and {Ibar}, Edo and {Inami}, Hanae and {Infante}, Leopoldo and {Hodge}, Jacqueline and {Karim}, Alex and {Le Fevre}, Olivier and {Magnelli}, Benjamin and {Neri}, Roberto and {Oesch}, Pascal and {Ota}, Kazuaki and {Popping}, Gerg{\"o} and {Rix}, Hans-Walter and {Sargent}, Mark and {Sheth}, Kartik and {van der Wel}, Arjen and {van der Werf}, Paul and {Wagg}, Jeff},
        title = "{ALMA Spectroscopic Survey in the Hubble Ultra Deep Field: Survey Description}",
      journal = {\apj},
         year = 2016,
        month = dec,
       volume = {833},
       number = {1},
          eid = {67},
        pages = {67},
          doi = {10.3847/1538-4357/833/1/67},
archivePrefix = {arXiv},
       eprint = {1607.06768},
 primaryClass = {astro-ph.GA},
       adsurl = {https://ui.adsabs.harvard.edu/abs/2016ApJ...833...67W}
}

@ARTICLE{Magnelli20,
       author = {{Magnelli}, Benjamin and {Boogaard}, Leindert and {Decarli}, Roberto and {G{\'o}nzalez-L{\'o}pez}, Jorge and {Novak}, Mladen and {Popping}, Gerg{\"o} and {Smail}, Ian and {Walter}, Fabian and {Aravena}, Manuel and {Assef}, Roberto J. and {Bauer}, Franz Erik and {Bertoldi}, Frank and {Carilli}, Chris and {Cortes}, Paulo C. and {Cunha}, Elisabete da and {Daddi}, Emanuele and {D{\'\i}az-Santos}, Tanio and {Inami}, Hanae and {Ivison}, Robert J. and {F{\`e}vre}, Olivier Le and {Oesch}, Pascal and {Riechers}, Dominik and {Rix}, Hans-Walter and {Sargent}, Mark T. and {Werf}, Paul van der and {Wagg}, Jeff and {Weiss}, Axel},
        title = "{The ALMA Spectroscopic Survey in the HUDF: The Cosmic Dust and Gas Mass Densities in Galaxies up to z {\ensuremath{\sim}} 3}",
      journal = {\apj},
         year = 2020,
        month = mar,
       volume = {892},
       number = {1},
          eid = {66},
        pages = {66},
          doi = {10.3847/1538-4357/ab7897},
archivePrefix = {arXiv},
       eprint = {2002.08640},
 primaryClass = {astro-ph.GA},
       adsurl = {https://ui.adsabs.harvard.edu/abs/2020ApJ...892...66M}
}

@ARTICLE{Hatsukade16,
       author = {{Hatsukade}, Bunyo and {Kohno}, Kotaro and {Umehata}, Hideki and {Aretxaga}, Itziar and {Caputi}, Karina I. and {Dunlop}, James S. and {Ikarashi}, Soh and {Iono}, Daisuke and {Ivison}, Rob J. and {Lee}, Minju and {Makiya}, Ryu and {Matsuda}, Yuichi and {Motohara}, Kentaro and {Nakanishi}, Kouichiro and {Ohta}, Kouji and {Tadaki}, Ken-ich and {Tamura}, Yoichi and {Wang}, Wei-Hao and {Wilson}, Grant W. and {Yamaguchi}, Yuki and {Yun}, Min S.},
        title = "{SXDF-ALMA 2-arcmin$^{2}$ deep survey: 1.1-mm number counts}",
      journal = {\pasj},
         year = 2016,
        month = jun,
       volume = {68},
       number = {3},
          eid = {36},
        pages = {36},
          doi = {10.1093/pasj/psw026},
archivePrefix = {arXiv},
       eprint = {1602.08167},
 primaryClass = {astro-ph.GA},
       adsurl = {https://ui.adsabs.harvard.edu/abs/2016PASJ...68...36H}
}

@ARTICLE{Franco18,
       author = {{Franco}, M. and {Elbaz}, D. and {B{\'e}thermin}, M. and {Magnelli}, B. and {Schreiber}, C. and {Ciesla}, L. and {Dickinson}, M. and {Nagar}, N. and {Silverman}, J. and {Daddi}, E. and {Alexander}, D.~M. and {Wang}, T. and {Pannella}, M. and {Le Floc'h}, E. and {Pope}, A. and {Giavalisco}, M. and {Maury}, A.~J. and {Bournaud}, F. and {Chary}, R. and {Demarco}, R. and {Ferguson}, H. and {Finkelstein}, S.~L. and {Inami}, H. and {Iono}, D. and {Juneau}, S. and {Lagache}, G. and {Leiton}, R. and {Lin}, L. and {Magdis}, G. and {Messias}, H. and {Motohara}, K. and {Mullaney}, J. and {Okumura}, K. and {Papovich}, C. and {Pforr}, J. and {Rujopakarn}, W. and {Sargent}, M. and {Shu}, X. and {Zhou}, L.},
        title = "{GOODS-ALMA: 1.1 mm galaxy survey. I. Source catalog and optically dark galaxies}",
      journal = {\aap},
         year = 2018,
        month = dec,
       volume = {620},
          eid = {A152},
        pages = {A152},
          doi = {10.1051/0004-6361/201832928},
archivePrefix = {arXiv},
       eprint = {1803.00157},
 primaryClass = {astro-ph.GA},
       adsurl = {https://ui.adsabs.harvard.edu/abs/2018A&A...620A.152F}
}

@ARTICLE{Dunlop17,
       author = {{Dunlop}, J.~S. and {McLure}, R.~J. and {Biggs}, A.~D. and {Geach}, J.~E. and {Micha{\l}owski}, M.~J. and {Ivison}, R.~J. and {Rujopakarn}, W. and {van Kampen}, E. and {Kirkpatrick}, A. and {Pope}, A. and {Scott}, D. and {Swinbank}, A.~M. and {Targett}, T.~A. and {Aretxaga}, I. and {Austermann}, J.~E. and {Best}, P.~N. and {Bruce}, V.~A. and {Chapin}, E.~L. and {Charlot}, S. and {Cirasuolo}, M. and {Coppin}, K. and {Ellis}, R.~S. and {Finkelstein}, S.~L. and {Hayward}, C.~C. and {Hughes}, D.~H. and {Ibar}, E. and {Jagannathan}, P. and {Khochfar}, S. and {Koprowski}, M.~P. and {Narayanan}, D. and {Nyland}, K. and {Papovich}, C. and {Peacock}, J.~A. and {Rieke}, G.~H. and {Robertson}, B. and {Vernstrom}, T. and {Werf}, P.~P. van der and {Wilson}, G.~W. and {Yun}, M.},
        title = "{A deep ALMA image of the Hubble Ultra Deep Field}",
      journal = {\mnras},
         year = 2017,
        month = apr,
       volume = {466},
       number = {1},
        pages = {861-883},
          doi = {10.1093/mnras/stw3088},
archivePrefix = {arXiv},
       eprint = {1606.00227},
 primaryClass = {astro-ph.GA},
       adsurl = {https://ui.adsabs.harvard.edu/abs/2017MNRAS.466..861D}
}

@ARTICLE{Planck20,
       author = {{Planck Collaboration} and {Aghanim}, N. and {Akrami}, Y. and {Ashdown}, M. and {Aumont}, J. and {Baccigalupi}, C. and {Ballardini}, M. and {Banday}, A.~J. and {Barreiro}, R.~B. and {Bartolo}, N. and {Basak}, S. and {Battye}, R. and {Benabed}, K. and {Bernard}, J. -P. and {Bersanelli}, M. and {Bielewicz}, P. and {Bock}, J.~J. and {Bond}, J.~R. and {Borrill}, J. and {Bouchet}, F.~R. and {Boulanger}, F. and {Bucher}, M. and {Burigana}, C. and {Butler}, R.~C. and {Calabrese}, E. and {Cardoso}, J. -F. and {Carron}, J. and {Challinor}, A. and {Chiang}, H.~C. and {Chluba}, J. and {Colombo}, L.~P.~L. and {Combet}, C. and {Contreras}, D. and {Crill}, B.~P. and {Cuttaia}, F. and {de Bernardis}, P. and {de Zotti}, G. and {Delabrouille}, J. and {Delouis}, J. -M. and {Di Valentino}, E. and {Diego}, J.~M. and {Dor{\'e}}, O. and {Douspis}, M. and {Ducout}, A. and {Dupac}, X. and {Dusini}, S. and {Efstathiou}, G. and {Elsner}, F. and {En{\ss}lin}, T.~A. and {Eriksen}, H.~K. and {Fantaye}, Y. and {Farhang}, M. and {Fergusson}, J. and {Fernandez-Cobos}, R. and {Finelli}, F. and {Forastieri}, F. and {Frailis}, M. and {Fraisse}, A.~A. and {Franceschi}, E. and {Frolov}, A. and {Galeotta}, S. and {Galli}, S. and {Ganga}, K. and {G{\'e}nova-Santos}, R.~T. and {Gerbino}, M. and {Ghosh}, T. and {Gonz{\'a}lez-Nuevo}, J. and {G{\'o}rski}, K.~M. and {Gratton}, S. and {Gruppuso}, A. and {Gudmundsson}, J.~E. and {Hamann}, J. and {Handley}, W. and {Hansen}, F.~K. and {Herranz}, D. and {Hildebrandt}, S.~R. and {Hivon}, E. and {Huang}, Z. and {Jaffe}, A.~H. and {Jones}, W.~C. and {Karakci}, A. and {Keih{\"a}nen}, E. and {Keskitalo}, R. and {Kiiveri}, K. and {Kim}, J. and {Kisner}, T.~S. and {Knox}, L. and {Krachmalnicoff}, N. and {Kunz}, M. and {Kurki-Suonio}, H. and {Lagache}, G. and {Lamarre}, J. -M. and {Lasenby}, A. and {Lattanzi}, M. and {Lawrence}, C.~R. and {Le Jeune}, M. and {Lemos}, P. and {Lesgourgues}, J. and {Levrier}, F. and {Lewis}, A. and {Liguori}, M. and {Lilje}, P.~B. and {Lilley}, M. and {Lindholm}, V. and {L{\'o}pez-Caniego}, M. and {Lubin}, P.~M. and {Ma}, Y. -Z. and {Mac{\'\i}as-P{\'e}rez}, J.~F. and {Maggio}, G. and {Maino}, D. and {Mandolesi}, N. and {Mangilli}, A. and {Marcos-Caballero}, A. and {Maris}, M. and {Martin}, P.~G. and {Martinelli}, M. and {Mart{\'\i}nez-Gonz{\'a}lez}, E. and {Matarrese}, S. and {Mauri}, N. and {McEwen}, J.~D. and {Meinhold}, P.~R. and {Melchiorri}, A. and {Mennella}, A. and {Migliaccio}, M. and {Millea}, M. and {Mitra}, S. and {Miville-Desch{\^e}nes}, M. -A. and {Molinari}, D. and {Montier}, L. and {Morgante}, G. and {Moss}, A. and {Natoli}, P. and {N{\o}rgaard-Nielsen}, H.~U. and {Pagano}, L. and {Paoletti}, D. and {Partridge}, B. and {Patanchon}, G. and {Peiris}, H.~V. and {Perrotta}, F. and {Pettorino}, V. and {Piacentini}, F. and {Polastri}, L. and {Polenta}, G. and {Puget}, J. -L. and {Rachen}, J.~P. and {Reinecke}, M. and {Remazeilles}, M. and {Renzi}, A. and {Rocha}, G. and {Rosset}, C. and {Roudier}, G. and {Rubi{\~n}o-Mart{\'\i}n}, J.~A. and {Ruiz-Granados}, B. and {Salvati}, L. and {Sandri}, M. and {Savelainen}, M. and {Scott}, D. and {Shellard}, E.~P.~S. and {Sirignano}, C. and {Sirri}, G. and {Spencer}, L.~D. and {Sunyaev}, R. and {Suur-Uski}, A. -S. and {Tauber}, J.~A. and {Tavagnacco}, D. and {Tenti}, M. and {Toffolatti}, L. and {Tomasi}, M. and {Trombetti}, T. and {Valenziano}, L. and {Valiviita}, J. and {Van Tent}, B. and {Vibert}, L. and {Vielva}, P. and {Villa}, F. and {Vittorio}, N. and {Wandelt}, B.~D. and {Wehus}, I.~K. and {White}, M. and {White}, S.~D.~M. and {Zacchei}, A. and {Zonca}, A.},
        title = "{Planck 2018 results. VI. Cosmological parameters}",
      journal = {\aap},
         year = 2020,
        month = sep,
       volume = {641},
          eid = {A6},
        pages = {A6},
          doi = {10.1051/0004-6361/201833910},
archivePrefix = {arXiv},
       eprint = {1807.06209},
 primaryClass = {astro-ph.CO},
       adsurl = {https://ui.adsabs.harvard.edu/abs/2020A&A...641A...6P}
}

@INPROCEEDINGS{Petry24,
       author = {{Petry}, Dirk and {D{\'\i}az Trigo}, Mar{\'\i}a. and {Kneissl}, R{\"u}diger and {Toledo}, Ignacio and {Miyazaki}, Atsushi and {Takagi}, Toshinobu and {Barnes}, Ashley and {Bonanomi}, Francesca},
        title = "{New methods for ALMA angular-scale-based observation scheduling, quality assessment, and beam shaping II: refinements}",
    booktitle = {Proc. SPIE Astronomical Telescopes and Instrumentation},
         year = 2024,
       editor = {{Benn}, Chris R. and {Chrysostomou}, Antonio and {Storrie-Lombardi}, Lisa J.},
       series = {SPIE Conference Series},
       volume = {13098},
        month = jul,
          eid = {130980P},
        pages = {130980P},
          doi = {10.1117/12.3012662},
       adsurl = {https://ui.adsabs.harvard.edu/abs/2024SPIE13098E..0PP}
}

@ARTICLE{Boogaard20,
       author = {{Boogaard}, Leindert A. and {van der Werf}, Paul and {Weiss}, Axel and {Popping}, Gerg{\"o} and {Decarli}, Roberto and {Walter}, Fabian and {Aravena}, Manuel and {Bouwens}, Rychard and {Riechers}, Dominik and {Gonz{\'a}lez-L{\'o}pez}, Jorge and {Smail}, Ian and {Carilli}, Chris and {Kaasinen}, Melanie and {Daddi}, Emanuele and {Cox}, Pierre and {D{\'\i}az-Santos}, Tanio and {Inami}, Hanae and {Cortes}, Paulo C. and {Wagg}, Jeff},
        title = "{The ALMA Spectroscopic Survey in the Hubble Ultra Deep Field: CO Excitation and Atomic Carbon in Star-forming Galaxies at z = 1-3}",
      journal = {\apj},
         year = 2020,
        month = oct,
       volume = {902},
       number = {2},
          eid = {109},
        pages = {109},
          doi = {10.3847/1538-4357/abb82f},
archivePrefix = {arXiv},
       eprint = {2009.04348},
 primaryClass = {astro-ph.GA},
       adsurl = {https://ui.adsabs.harvard.edu/abs/2020ApJ...902..109B}
}

@ARTICLE{Dunne22,
       author = {{Dunne}, L. and {Maddox}, S.~J. and {Papadopoulos}, P.~P. and {Ivison}, R.~J. and {Gomez}, H.~L.},
        title = "{Dust, CO, and [C I]: cross-calibration of molecular gas mass tracers in metal-rich galaxies across cosmic time}",
      journal = {\mnras},
         year = 2022,
        month = nov,
       volume = {517},
       number = {1},
        pages = {962-999},
          doi = {10.1093/mnras/stac2098},
archivePrefix = {arXiv},
       eprint = {2208.01622},
 primaryClass = {astro-ph.GA},
       adsurl = {https://ui.adsabs.harvard.edu/abs/2022MNRAS.517..962D}
}

@PHDTHESIS{Briggs95,
       author = {{Briggs}, Daniel Shenon},
        title = "{High fidelity deconvolution of moderately resolved sources}",
       school = {New Mexico Institute of Mining and Technology},
         year = 1995,
        month = jan,
       adsurl = {https://ui.adsabs.harvard.edu/abs/1995PhDT.......238B}
}

@book{cortes2025, 
	author = "Cortes, P. C. and Remijan, A. and Hales, A. and Carpenter, J. and Dent, W. and Kameno, S. and Loomis, R. and Vila-Vilaro, B. and Biggs, A. and Miotello, A. and Vlahakis, C. and Rosen, R. and Stoehr, F. and Saini, K.",
	publisher = {ALMA Doc 12.3, version 1.0},
	title  = "ALMA Technical Handbook Cycle 12", 
	year   = 2025,
        month  = mar,
	note = "https://almascience.eso.org/documents-and-tools/cycle12/alma-technical-handbook"}

@book{cortes2022, 
	author = "Cortes, P. C. and Remijan, A. and Hales, A. and Carpenter, J. and Dent, W. and Kameno, S. and Loomis, R. and Vila-Vilaro, B. and Biggs, A. and Miotello, A. and Vlahakis, C. and Rosen, R. and Stoehr, F. and Saini, K.",
	publisher = {ALMA Doc. 9.3, version 1.0},
	title  = "ALMA Technical Handbook Cycle 9", 
	year   = 2022,
	note = "https://almascience.eso.org/documents-and-tools/cycle9/alma-technical-handbook"}

@ARTICLE{Verhamme18,
       author = {{Verhamme}, A. and {Garel}, T. and {Ventou}, E. and {Contini}, T. and {Bouch{\'e}}, N. and {Herenz}, EC and {Richard}, J. and {Bacon}, R. and {Schmidt}, KB and {Maseda}, M. and {Marino}, RA and {Brinchmann}, J. and {Cantalupo}, S. and {Caruana}, J. and {Cl{\'e}ment}, B. and {Diener}, C. and {Drake}, AB and {Hashimoto}, T. and {Inami}, H. and {Kerutt}, J. and {Kollatschny}, W. and {Leclercq}, F. and {Patr{\'\i}cio}, V. and {Schaye}, J. and {Wisotzki}, L. and {Zabl}, J.},
        title = "{Recovering the systemic redshift of galaxies from their Lyman alpha line profile}",
      journal = {\mnras},
         year = 2018,
        month = jul,
       volume = {478},
       number = {1},
        pages = {L60-L65},
          doi = {10.1093/mnrasl/sly058},
archivePrefix = {arXiv},
       eprint = {1804.01883},
 primaryClass = {astro-ph.GA},
       adsurl = {https://ui.adsabs.harvard.edu/abs/2018MNRAS.478L..60V}
}

@ARTICLE{Ferguson00,
       author = {{Ferguson}, Henry C. and {Dickinson}, Mark and {Williams}, Robert},
        title = "{The Hubble Deep Fields}",
      journal = {\araa},
         year = 2000,
        month = jan,
       volume = {38},
        pages = {667-715},
          doi = {10.1146/annurev.astro.38.1.667},
archivePrefix = {arXiv},
       eprint = {astro-ph/0004319},
 primaryClass = {astro-ph},
       adsurl = {https://ui.adsabs.harvard.edu/abs/2000ARA&A..38..667F}
}

@ARTICLE{Bacon23,
       author = {{Bacon}, Roland and {Brinchmann}, Jarle and {Conseil}, Simon and {Maseda}, Michael and {Nanayakkara}, Themiya and {Wendt}, Martin and {Bacher}, Raphael and {Mary}, David and {Weilbacher}, Peter M. and {Krajnovi{\'c}}, Davor and {Boogaard}, Leindert and {Bouch{\'e}}, Nicolas and {Contini}, Thierry and {Epinat}, Beno{\^\i}t and {Feltre}, Anna and {Guo}, Yucheng and {Herenz}, Christian and {Kollatschny}, Wolfram and {Kusakabe}, Haruka and {Leclercq}, Floriane and {Michel-Dansac}, L{\'e}o and {Pello}, Roser and {Richard}, Johan and {Roth}, Martin and {Salvignol}, Gregory and {Schaye}, Joop and {Steinmetz}, Matthias and {Tresse}, Laurence and {Urrutia}, Tanya and {Verhamme}, Anne and {Vitte}, Eloise and {Wisotzki}, Lutz and {Zoutendijk}, Sebastiaan L.},
        title = "{The MUSE Hubble Ultra Deep Field surveys: Data release II}",
      journal = {\aap},
         year = 2023,
        month = feb,
       volume = {670},
          eid = {A4},
        pages = {A4},
          doi = {10.1051/0004-6361/202244187},
archivePrefix = {arXiv},
       eprint = {2211.08493},
 primaryClass = {astro-ph.GA},
       adsurl = {https://ui.adsabs.harvard.edu/abs/2023A&A...670A...4B}
}

@ARTICLE{Westmeier21,
       author = {{Westmeier}, T. and {Kitaeff}, S. and {Pallot}, D. and {Serra}, P. and {van der Hulst}, J.~M. and {Jurek}, R.~J. and {Elagali}, A. and {For}, B. -Q. and {Kleiner}, D. and {Koribalski}, B.~S. and {Lee-Waddell}, K. and {Mould}, J.~R. and {Reynolds}, T.~N. and {Rhee}, J. and {Staveley-Smith}, L.},
        title = "{SOFIA 2 - an automated, parallel H I source finding pipeline for the WALLABY survey}",
      journal = {\mnras},
         year = 2021,
        month = sep,
       volume = {506},
       number = {3},
        pages = {3962-3976},
          doi = {10.1093/mnras/stab1881},
archivePrefix = {arXiv},
       eprint = {2106.15789},
 primaryClass = {astro-ph.IM},
       adsurl = {https://ui.adsabs.harvard.edu/abs/2021MNRAS.506.3962W}
}

@ARTICLE{Serra15,
       author = {{Serra}, Paolo and {Westmeier}, Tobias and {Giese}, Nadine and {Jurek}, Russell and {Fl{\"o}er}, Lars and {Popping}, Attila and {Winkel}, Benjamin and {van der Hulst}, Thijs and {Meyer}, Martin and {Koribalski}, B{\"a}rbel S. and {Staveley-Smith}, Lister and {Courtois}, H{\'e}l{\`e}ne},
        title = "{SOFIA: a flexible source finder for 3D spectral line data}",
      journal = {\mnras},
         year = 2015,
        month = apr,
       volume = {448},
       number = {2},
        pages = {1922-1929},
          doi = {10.1093/mnras/stv079},
archivePrefix = {arXiv},
       eprint = {1501.03906},
 primaryClass = {astro-ph.IM},
       adsurl = {https://ui.adsabs.harvard.edu/abs/2015MNRAS.448.1922S}
}

@ARTICLE{Bacon15,
       author = {{Bacon}, R. and {Brinchmann}, J. and {Richard}, J. and {Contini}, T. and {Drake}, A. and {Franx}, M. and {Tacchella}, S. and {Vernet}, J. and {Wisotzki}, L. and {Blaizot}, J. and {Bouch{\'e}}, N. and {Bouwens}, R. and {Cantalupo}, S. and {Carollo}, C.~M. and {Carton}, D. and {Caruana}, J. and {Cl{\'e}ment}, B. and {Dreizler}, S. and {Epinat}, B. and {Guiderdoni}, B. and {Herenz}, C. and {Husser}, T. -O. and {Kamann}, S. and {Kerutt}, J. and {Kollatschny}, W. and {Krajnovic}, D. and {Lilly}, S. and {Martinsson}, T. and {Michel-Dansac}, L. and {Patricio}, V. and {Schaye}, J. and {Shirazi}, M. and {Soto}, K. and {Soucail}, G. and {Steinmetz}, M. and {Urrutia}, T. and {Weilbacher}, P. and {de Zeeuw}, T.},
        title = "{The MUSE 3D view of the Hubble Deep Field South}",
      journal = {\aap},
         year = 2015,
        month = mar,
       volume = {575},
          eid = {A75},
        pages = {A75},
          doi = {10.1051/0004-6361/201425419},
archivePrefix = {arXiv},
       eprint = {1411.7667},
 primaryClass = {astro-ph.GA},
       adsurl = {https://ui.adsabs.harvard.edu/abs/2015A&A...575A..75B}
}

@article{Hunter23,
doi = {10.1088/1538-3873/ace216},
url = {https://dx.doi.org/10.1088/1538-3873/ace216},
year = 2023,
month = jul,
publisher = {The Astronomical Society of the Pacific},
volume = {135},
number = {1049},
pages = {074501},
author = {Todd R. Hunter and Remy Indebetouw and Crystal L. Brogan and Kristin Berry and Chin-Shin Chang and Harold Francke and Vincent C. Geers and Laura Gomez and John E. Hibbard and Elizabeth M. Humphreys and Brian R. Kent and Amanda A. Kepley and Devaky Kunneriath and Andrew Lipnicky and Ryan A. Loomis and Brian S. Mason and Joseph S. Masters and Luke T. Maud and Dirk Muders and Jose Sabater and Kanako Sugimoto and LÃ¡szlÃ³ SzÅ±cs and Eugene Vasiliev and Liza Videla and Eric Villard and Stewart J. Williams and Rui Xue and Ilsang Yoon},
title = {The ALMA Interferometric Pipeline Heuristics},
journal = {Publications of the Astronomical Society of the Pacific},
}

@ARTICLE{casa2022,
       author = {{CASA Team} and {Bean}, Ben and {Bhatnagar}, Sanjay and {Castro}, Sandra and {Donovan Meyer}, Jennifer and {Emonts}, Bjorn and {Garcia}, Enrique and {Garwood}, Robert and {Golap}, Kumar and {Gonzalez Villalba}, Justo and {Harris}, Pamela and {Hayashi}, Yohei and {Hoskins}, Josh and {Hsieh}, Mingyu and {Jagannathan}, Preshanth and {Kawasaki}, Wataru and {Keimpema}, Aard and {Kettenis}, Mark and {Lopez}, Jorge and {Marvil}, Joshua and {Masters}, Joseph and {McNichols}, Andrew and {Mehringer}, David and {Miel}, Renaud and {Moellenbrock}, George and {Montesino}, Federico and {Nakazato}, Takeshi and {Ott}, Juergen and {Petry}, Dirk and {Pokorny}, Martin and {Raba}, Ryan and {Rau}, Urvashi and {Schiebel}, Darrell and {Schweighart}, Neal and {Sekhar}, Srikrishna and {Shimada}, Kazuhiko and {Small}, Des and {Steeb}, Jan-Willem and {Sugimoto}, Kanako and {Suoranta}, Ville and {Tsutsumi}, Takahiro and {van Bemmel}, Ilse M. and {Verkouter}, Marjolein and {Wells}, Akeem and {Xiong}, Wei and {Szomoru}, Arpad and {Griffith}, Morgan and {Glendenning}, Brian and {Kern}, Jeff},
        title = "{CASA, the Common Astronomy Software Applications for Radio Astronomy}",
      journal = {\pasp},
         year = 2022,
        month = nov,
       volume = {134},
       number = {1041},
          eid = {114501},
        pages = {114501},
          doi = {10.1088/1538-3873/ac9642},
archivePrefix = {arXiv},
       eprint = {2210.02276},
 primaryClass = {astro-ph.IM},
       adsurl = {https://ui.adsabs.harvard.edu/abs/2022PASP..134k4501C}
}

\begin{appendix} 

\section{Observation details} \label{app:ebs}

Table~\ref{tab:ebs} lists the 70 execution blocks (EBs) used to build the continuum maps and cubes. The columns contain: EB Unique IDentifiers (UIDs), date of observation, number of antennas in the array, configuration, minimum and maximum baseline, J-name of the Bandpass calibrator (also used for flux calibration), Phase calibrator, and time on science target.

\clearpage
\onecolumn
\begin{landscape}
\begin{longtable}{l l c c c c c c c}
\caption{Details of the observed execution blocks.}
\label{tab:ebs}\\
\hline\hline
EB & Date	&	Number of	&	Approx &	\multicolumn{2}{c}{Baseline}	&	Amplitude, Bandpass &	Phase &	Time on source	\\
    &  &   antennas & Config & min [m]  &  max [km] & calibrator & calibrator & [mm:ss] \\
\hline
\endfirsthead
\caption{Details of the observed execution blocks (continued).}\\
\hline\hline
EB & Date	&	Number of	&	Approx &	\multicolumn{2}{c}{Baseline}	&	Amplitude, Bandpass &	Phase &	Time on source	\\
    &  &   antennas & Config & min [m]  &  max [km] & calibrator & calibrator & [mm:ss] \\
\hline
\endhead
uid://A002/X10c5c57/X2e6b	& 2023-09-05 &	46	&	C-9	&	83.1 &	14.9 &	J2357-5311 	&	J2239-5701 	&	53:24	\\
uid://A002/X10c6fac/X341d	&	2023-09-07	&	46	&	C-9	&	83.1	&	14.9	&	J2357-5311 	&	J2239-5701 	&	53:30	\\
uid://A002/X10c6fac/X8a9e	&	2023-09-08	&	46	&	C-9	&	83.1	&	14.9	&	J2258-2758 	&	J2239-5701 	&	53:24	\\
uid://A002/X10ca5ad/X7b50	&	2023-09-12	&	49	&	C-8	&	66.8	&	11.9	&	J2357-5311 	&	J2239-5701 	&	10:25	\\
uid://A002/X10ca5ad/Xb62f	&	2023-09-13	&	47	&	C-8	&	66.8	&	11.9	&	J2357-5311 	&	J2239-5701 	&	53:24	\\
uid://A002/X10cc13c/Xe4a	&	2023-09-14	&	45	&	C-8	&	83.1	&	11.9	&	J2357-5311 	&	J2239-5701 	&	53:24	\\
uid://A002/X10cc13c/X1c61	&	2023-09-14	&	45	&	C-8	&	83.1	&	11.9	&	J2357-5311 	&	J2239-5701 	&	53:24	\\
uid://A002/X10cc13c/Xa4fb	&	2023-09-15	&	46	&	C-8	&	83.1	&	9.7	&	J2357-5311 	&	J2239-5701 	&	53:24	\\
uid://A002/X10cc13c/Xb109	&	2023-09-15	&	45	&	C-8	&	83.1	&	9.7	&	J2357-5311 	&	J2239-5701 	&	53:24	\\
uid://A002/X10cc13c/Xb95d	&	2023-09-15	&	48	&	C-8	&	83.1	&	9.7	&	J2357-5311 	&	J2239-5701 	&	53:24	\\
uid://A002/X10cc13c/X121fd	&	2023-09-16	&	46	&	C-8	&	92.1	&	8.8	&	J2357-5311 	&	J2239-5701 	&	53:24	\\
uid://A002/X10ce151/X26a4	&	2023-09-17	&	41	&	C-8	&	92.1	&	8.3	&	J2357-5311 	&	J2239-5701 	&	53:24	\\
uid://A002/X10ceb22/X1990	&	2023-09-18	&	43	&	C-8	&	92.1	&	8.3	&	J2357-5311 	&	J2239-5701 	&	53:24	\\
uid://A002/X10ceb22/X2d5e	&	2023-09-18	&	43	&	C-8	&	92.1	&	8.5	&	J2357-5311 	&	J2239-5701 	&	53:26	\\
uid://A002/X10ceb22/Xa64c	&	2023-09-19	&	41	&	C-8	&	97.1	&	8.3	&	J2357-5311 	&	J2239-5701 	&	53:24	\\
uid://A002/X10ceb22/Xaf86	&	2023-09-19	&	45	&	C-8	&	92.1	&	8.5	&	J2357-5311 	&	J2239-5701 	&	53:24	\\
uid://A002/X10ceb22/X11957	&	2023-09-20	&	46	&	C-8	&	92.1	&	8.5	&	J2357-5311 	&	J2239-5701 	&	53:24	\\
uid://A002/X10d12a2/X3ff	&	2023-09-21	&	43	&	C-8	&	92.1	&	8.5	&	J2357-5311 	&	J2239-5701 	&	53:24	\\
uid://A002/X10d12a2/X6b30	&	2023-09-22	&	44	&	C-8	&	92.1	&	8.5	&	J2357-5311 	&	J2239-5701 	&	53:24	\\
uid://A002/X10d12a2/X70b8	&	2023-09-22	&	44	&	C-8	&	92.1	&	8.5	&	J2357-5311 	&	J2239-5701 	&	53:24	\\
uid://A002/X10d12a2/Xe1a2	&	2023-09-23	&	44	&	C-8	&	92.1	&	8.3	&	J2357-5311 	&	J2239-5701 	&	53:25	\\
uid://A002/X10d12a2/Xed89	&	2023-09-23	&	45	&	C-8	&	89.6	&	8.3	&	J2357-5311 	&	J2239-5701 	&	53:24	\\
uid://A002/X10d4e2e/X36ee	&	2023-09-26	&	45	&	C-8	&	92.1	&	8.5	&	J2357-5311 	&	J2239-5701 	&	53:24	\\
uid://A002/X10d4e2e/X4034	&	2023-09-26	&	45	&	C-8	&	92.1	&	8.5	&	J2357-5311 	&	J2239-5701 	&	53:24	\\
uid://A002/X10d4e2e/X4740	&	2023-09-26	&	45	&	C-8	&	92.1	&	8.5	&	J2357-5311 	&	J2239-5701 	&	53:24	\\
uid://A002/X10d4e2e/Xc316	&	2023-09-27	&	46	&	C-8	&	92.1	&	8.5	&	J2357-5311 	&	J2239-5701 	&	53:24	\\
uid://A002/X10d4e2e/X109f6	&	2023-09-28	&	47	&	C-8	&	92.1	&	8.5	&	J2357-5311 	&	J2239-5701 	&	53:24	\\
uid://A002/X10dc577/X2e95	&	2023-10-07	&	44	&	C-8	&	92.1	&	8.5	&	J2357-5311 	&	J2239-5701 	&	53:24	\\
uid://A002/X10dc577/X42bc	&	2023-10-07	&	44	&	C-8	&	92.1	&	8.5	&	J2357-5311 	&	J2239-5701 	&	53:24	\\
uid://A002/X10e09b4/X4b1	&	2023-10-11	&	42	&	C-8	&	92.1	&	8.3	&	J2357-5311 	&	J2239-5701 	&	53:24	\\
uid://A002/X10e09b4/Xa9f2	&	2023-10-13	&	45	&	C-8	&	92.1	&	8.5	&	J2357-5311 	&	J2239-5701 	&	53:24	\\
uid://A002/X10e2702/X597	&	2023-10-13	&	43	&	C-8	&	92.1	&	8.3	&	J2357-5311 	&	J2239-5701 	&	53:24	\\
uid://A002/X10e2702/X2b89	&	2023-10-14	&	44	&	C-8	&	92.1	&	8.3	&	J2357-5311 	&	J2239-5701 	&	16:17	\\
uid://A002/X10e318c/X1b4b	&	2023-10-15	&	41	&	C-8	&	89.6	&	8.3	&	J2357-5311 	&	J2239-5701 	&	53:24	\\
uid://A002/X10e318c/X22cb	&	2023-10-15	&	41	&	C-8	&	89.6	&	8.3	&	J2357-5311 	&	J2239-5701 	&	53:24	\\
uid://A002/X10e3e4c/Xa19	&	2023-10-15	&	42	&	C-8	&	92.1	&	8.3	&	J2357-5311 	&	J2239-5701 	&	53:24	\\
uid://A002/X10e3e4c/X11d3	&	2023-10-16	&	42	&	C-8	&	92.1	&	8.3	&	J2357-5311 	&	J2239-5701 	&	53:24	\\
uid://A002/X10e3e4c/X1c6e	&	2023-10-16	&	41	&	C-8	&	92.1	&	8.3	&	J2357-5311 	&	J2239-5701 	&	53:24	\\
uid://A002/X10e492f/X35b3	&	2023-10-16	&	44	&	C-8	&	92.1	&	8.3	&	J2357-5311 	&	J2239-5701 	&	53:24	\\
uid://A002/X10e492f/X41b8	&	2023-10-17	&	44	&	C-8	&	92.1	&	8.3	&	J2357-5311 	&	J2239-5701 	&	53:24	\\
uid://A002/X10e492f/X4ffc	&	2023-10-17	&	44	&	C-8	&	92.1	&	8.3	&	J2357-5311 	&	J2239-5701 	&	53:24	\\
uid://A002/X10e492f/Xb325	&	2023-10-17	&	46	&	C-8	&	92.1	&	8.5	&	J2357-5311 	&	J2239-5701 	&	53:24	\\
uid://A002/X10e492f/Xc5f6	&	2023-10-18	&	43	&	C-8	&	92.1	&	8.3	&	J2357-5311 	&	J2239-5701 	&	53:24	\\
uid://A002/X10e6d25/X2460	&	2023-10-20	&	42	&	C-8	&	92.1	&	8.3	&	J2357-5311 	&	J2239-5701 	&	53:24	\\
uid://A002/X10e6d25/X352e	&	2023-10-20	&	45	&	C-8	&	92.1	&	8.3	&	J2357-5311 	&	J2239-5701 	&	53:24	\\
uid://A002/X10e6d25/X93f5	&	2023-10-20	&	45	&	C-8	&	92.1	&	8.3	&	J2357-5311 	&	J2239-5701 	&	53:24	\\
uid://A002/X10e6d25/Xa314	&	2023-10-21	&	46	&	C-8	&	92.1	&	8.3	&	J2357-5311 	&	J2239-5701 	&	53:24	\\
uid://A002/X10e8a11/X14a7	&	2023-10-21	&	47	&	C-8	&	92.1	&	8.5	&	J2357-5311 	&	J2239-5701 	&	53:24	\\
uid://A002/X10e8a11/X1f2d	&	2023-10-22	&	47	&	C-8	&	92.1	&	8.5	&	J2357-5311 	&	J2239-5701 	&	53:24	\\
uid://A002/X10e9b60/X162b	&	2023-10-24	&	43	&	C-8	&	92.1	&	8.3	&	J2357-5311 	&	J2239-5701 	&	53:24	\\
uid://A002/X10e9b60/X2419	&	2023-10-24	&	43	&	C-8	&	92.1	&	8.3	&	J2357-5311 	&	J2239-5701 	&	53:24	\\
uid://A002/X10e9b60/Xa45b	&	2023-10-25	&	45	&	C-8	&	92.1	&	8.3	&	J2357-5311 	&	J2239-5701 	&	45:55	\\
uid://A002/X10e9b60/Xb860	&	2023-10-25	&	44	&	C-8	&	113	&	8.3	&	J2357-5311 	&	J2239-5701 	&	53:24	\\
uid://A002/X10eced8/X139b	&	2023-10-29	&	49	&	C-8	&	66.8	&	8.3	&	J2357-5311 	&	J2239-5701 	&	53:24	\\
uid://A002/X10ed869/X2125	&	2023-10-30	&	45	&	C-8	&	89.6	&	8.3	&	J2357-5311 	&	J2239-5701 	&	53:24	\\
uid://A002/X10ed869/X15ed8	&	2023-11-02	&	43	&	C-7 hybrid	&	85.2	&	6.6	&	J2357-5311 	&	J2239-5701 	&	42:00	\\
uid://A002/X10ed869/X1bc53	&	2023-11-02	&	41	&	C-7 hybrid	&	85.2	&	8.3	&	J2357-5311 	&	J2239-5701 	&	53:24	\\
uid://A002/X10ed869/X1cb97	&	2023-11-03	&	41	&	C-7 hybrid	&	85.2	&	8.3	&	J2357-5311 	&	J2239-5701 	&	53:24	\\
uid://A002/X10f13dc/X44fa	&	2023-11-03	&	43	&	C-7 hybrid	&	85.2	&	6.6	&	J2357-5311 	&	J2239-5701 	&	53:24	\\
uid://A002/X10f262c/X4cb	&	2023-11-04	&	42	&	C-7 hybrid	&	30.9	&	6.6	&	J2357-5311 	&	J2239-5701 	&	53:24	\\
uid://A002/X10f262c/Xe18	&	2023-11-05	&	42	&	C-7 hybrid	&	30.9	&	6.6	&	J2357-5311 	&	J2239-5701 	&	53:24	\\
uid://A002/X10f262c/X179d	&	2023-11-05	&	42	&	C-7 hybrid	&	30.9	&	6.6	&	J2357-5311 	&	J2239-5701 	&	53:24	\\
uid://A002/X10f30e6/X561	&	2023-11-05	&	43	&	C-7 hybrid	&	85.2	&	6.6	&	J2357-5311 	&	J2239-5701 	&	53:24	\\
uid://A002/X10f30e6/X119d	&	2023-11-06	&	43	&	C-7 hybrid	&	85.2	&	6.6	&	J2357-5311 	&	J2239-5701 	&	53:24	\\
uid://A002/X10f3768/X2d99	&	2023-11-06	&	44	&	C-7 hybrid	&	85.2	&	6.6	&	J2357-5311 	&	J2239-5701 	&	53:24	\\
uid://A002/X10f3768/X3951	&	2023-11-06	&	44	&	C-7 hybrid	&	85.2	&	6.6	&	J2357-5311 	&	J2239-5701 	&	53:24	\\
uid://A002/X10f3768/Xf62d	&	2023-11-08	&	49	&	C-7 hybrid	&	85.2	&	6.6	&	J2357-5311 	&	J2239-5701 	&	53:24	\\
uid://A002/X10fdea7/X4b22	&	2023-11-23	&	42	&	C-7 hybrid	&	45	&	5.2	&	J2357-5311 	&	J2239-5701 	&	53:35	\\
uid://A002/X1103e51/X13a	&	2023-11-29	&	42	&	C-7	&	41.4	&	3.6	&	J2357-5311 	&	J2239-5701 	&	53:24	\\
uid://A002/X1104ab5/X2ed	&	2023-11-30	&	43	&	C-7	&	64.1	&	3.1	&	J2357-5311 	&	J2239-5701 	&	53:24	\\
\hline
\end{longtable}
\end{landscape}
\clearpage

\section{Comparison with near-IR emission} \label{app:irac} 

\begin{figure*}
    \centering
    \includegraphics[width=0.75\linewidth]{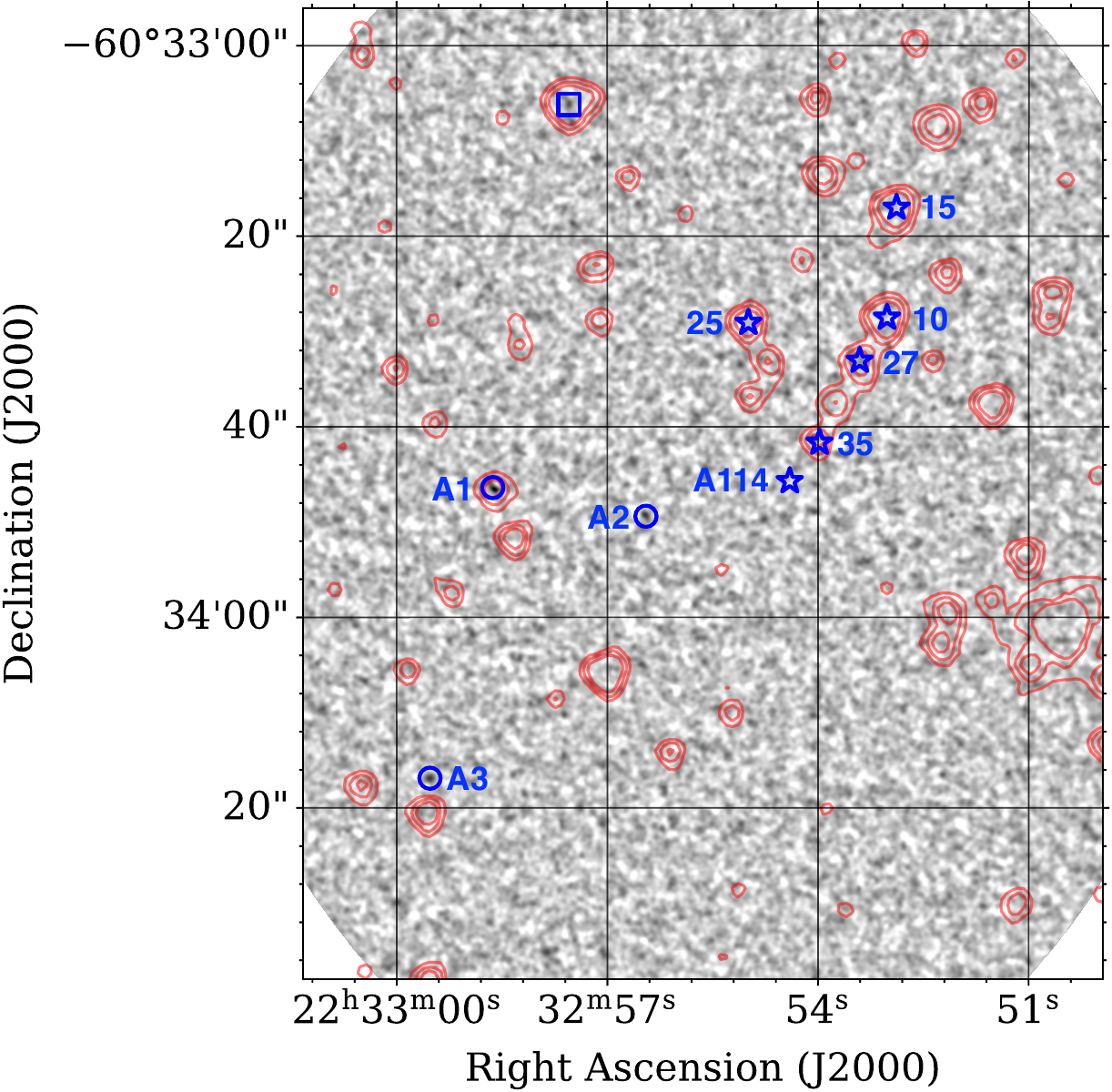}
    \caption{The natural-weighted 3mm-continuum map with a 0.5'' $uv$-taper (gray scale) with contours overlaid from the IRAC-3.6\,$\mu$m map (red isocontour levels at 3, 6, and 12$\sigma$). Sources which revealed CO~(2-1) line emission are marked with open blue stars, while the continuum detections reported in Table~\ref{tab:contdet} are marked as blue circled. The potential $3\sigma$ continuum detection toward an IRAC source is marked with a blue square.}
    \label{fig:irac}
\end{figure*}

In Figure~\ref{fig:irac}, we compare the ALMA 3\,mm continuum map (natural-weighted with a 0.5'' $uv$-taper as used for continuum source detection) with \textit{Spitzer}/IRAC-3.6\,$\mu$m map. Only one of the three reliable continuum detections has a match with an IRAC-detected source. Moreover, we note that we do find evidence for a $\sim3\sigma$ detection ($21.3\pm6.9\,\mu$Jy) towards an IRAC source at RA,~Dec\,=\,22:32:57.55,\,-60:33:06.2 (at a PB attenuation of 0.54). The five main line detections (Figures~\ref{fig:b1510} through \ref{fig:b1535}) are all detected in IRAC (rest-frame $\sim$1.6\,$\mu$m), while the serendipitously found line emitter ALMA-114 is not. Finally, it is worth noting that the filamentary structure, with B15-10 and ALMA-114 placed at its extremes ($\sim20$'' or $\sim170\,$kpc apart), is showing evidence of the type of environment these sources are found in.

\section{Comparison with rest-frame UV-optical emission} \label{app:hst} 

\begin{figure*}
    \centering
    \includegraphics[width=\linewidth]{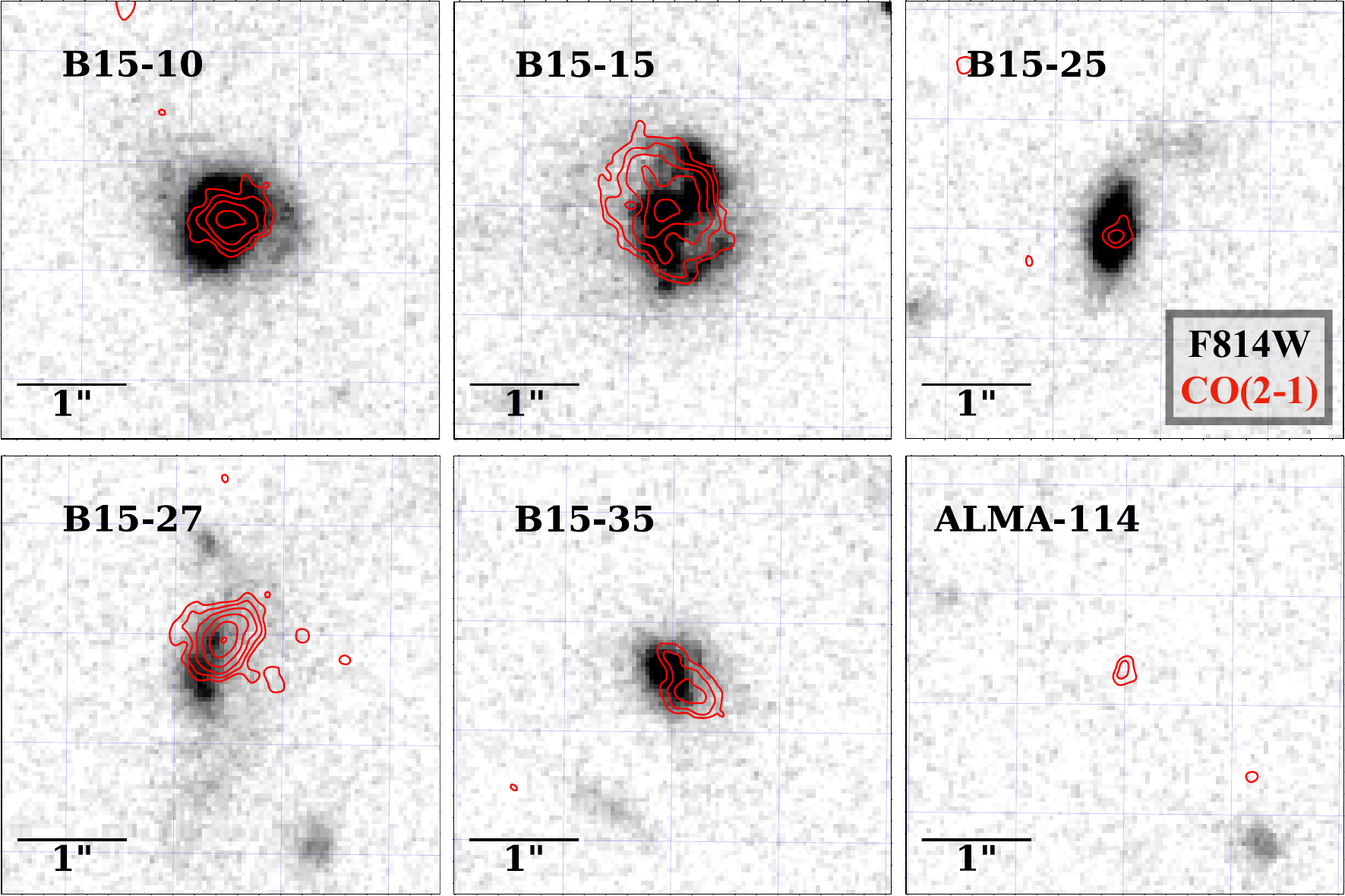}
    \caption{Comparison between \textit{HST} WFPC2/\textit{F814W} imaging (background gray image) and the CO\,(2-1) maps (red contours) for the detected sources. 
    Cutouts are 4\arcsec\ wide ($\sim34\,$kpc).
    The contours indicate levels at $3\times\sqrt(2)^{n}~\times~${\sc rms}, where $n=0,\,1,\,...$.
    B15-114 is detected by \textit{HST} in the South-West corner of the ALMA-114 panel, but not in CO\,(2-1). 
    }
    \label{fig:hst}
\end{figure*}

In this section we provide a comparison between the CO\,(2-1) emission from sources within the group at $z_{\rm spec}=1.284$ and the rest-frame UV-optical emission traced by \textit{HST} imaging. Specifically the longest-wavelength WFPC2 filter \textit{F814W} ($\lambda=827\pm88\,$nm) traces $\lambda_{\rm rest}=362\pm40\,$nm) at $z_{\rm spec}=1.284$. \textit{HST} imaging was aligned with \textit{Gaia}-DR3 \citep{Gaia23} making use of three stars in the field (details in Figure~\ref{fig:almaptg}). Figure~\ref{fig:hst} shows \textit{F814W} maps with CO\,(2-1) maps overlaid on the six sources with detected CO emission. One can clearly see the spatial offset between star-formation regions traced by HST and the molecular gas emission in B15-27 and B15-35. This is especially interesting since B15-27 is the source to show the cleanest rotation pattern in the sample (see more discussion in Section~\ref{sec:lines}). ALMA-114 is clearly undetected by HST, while its companion B15-114 is in the South-West region.

\end{appendix}

\end{document}